%% file: YbRh2Si2_vArXiv.tex
\documentclass[aps,prl,preprint,floatfix,showpacs]{revtex4-1}

\usepackage{bm}
\usepackage{xr}
\usepackage{epsfig}
\usepackage{comment}
\usepackage{epstopdf}
\usepackage{graphicx}
\usepackage{graphics}
\DeclareGraphicsExtensions{.eps}
\usepackage{color}

\usepackage{enumitem}

\usepackage{amsmath}
\usepackage{amsfonts}
\usepackage{amssymb}

\begin{document}

\title{Thermal conductivity through the quantum critical point in YbRh$_2$Si$_2$ at very low temperature}

\author{M.~Taupin$^1$} 
\thanks{Present address: Low Temperature Laboratory, Aalto University, P.O. Box 13500, FI-00076 Aalto, Finland}
\author{G.~Knebel$^1$}
\author{T.~D.~Matsuda$^{2,3}$}
\author{G. Lapertot$^1$}
\author{Y. Machida$^4$}
\author{K. Izawa$^4$}
\author{J.-P.~Brison$^1$}
\email[]{jean-pascal.brison@cea.fr}
\author{J.~Flouquet$^1$} 

\affiliation{$^1$ Univ. Grenoble Alpes, INAC-SPSMS, F-38000 Grenoble, France\\
$^{\ }$CEA, INAC-SPSMS, F-38000 Grenoble, France\\
$^2$ Advanced Science Research Center, JAEA, Tokai, Ibaraki 319-1195, Japan\\
$^3$ Department of Physics, Tokyo Metropolitan Univ. 1-1 Minami-Osawa, Hachioji-shi, Tokyo 192-0397, Japan\\
$^4$ Department of Physics, Tokyo Institute of Technology, Meguro 152-8551, Japan}

\date{\today}

\begin{abstract}
The thermal conductivity of YbRh$_2$Si$_2$ has been measured down to very low temperatures under field in the basal plane. An additional channel for heat transport appears below 30\,mK, both in the antiferromagnetic and paramagnetic states, respectively below and above the critical field suppressing the magnetic order. This excludes antiferromagnetic magnons as the origin of this additional contribution to thermal conductivity. Moreover, this low temperature contribution prevails a definite conclusion on the validity or violation of the Wiedemann-Franz law at the field-induced quantum critical point. At high temperature in the paramagnetic state, the thermal conductivity is sensitive to ferromagnetic fluctuations, previously observed by NMR or neutron scattering and required for the occurrence of the sharp electronic spin resonance fracture.
\end{abstract}



\maketitle

Traditionally, heavy fermion intermetallics are viewed as systems lying close to the instability to an ordered magnetic state. The instability is driven by the competition between Kondo screening of the local moments, and long range order of these moments, induced by RKKY interactions. Along the past decades, experimental and theoretical studies have endeavored to unveil the evolution of heavy fermions through the instability, focusing on the role of quantum fluctuations around the quantum critical point (QCP) \cite{Lohneysen2007}. Presently, the most prominent scenarios involve instabilities approaching the critical point from the paramagnetic (PM) side, either on some ``hot spots'', which is the so-called ``spin density wave'' scenario, or extended over the whole surface, for the ``local quantum critical scenario''. In the last case, the Kondo screening mechanism would breakdown near the quantum critical point, leading to a change from a large Fermi surface (including the local moments) in the paramagnetic state, to a ``small Fermi surface'' when Kondo coupling disappears \cite{Coleman2001}. As a consequence, it is also believed that in this last scenario, which involves a collapse of the mechanism driving the heavy quasi-particles formation, when approaching the well-defined QCP, the low energy excitations would not be described by quasiparticles anymore. So experiments probing this local quantum critical scenario, are mainly trying to reveal a change of Fermi surface (FS) volume across the QCP or a clear violation of predictions based on the notion of quasiparticles.

The antiferromagnetic (AF) heavy fermion YbRh$_2$Si$_2$, has been a main playground for both experimental approaches, and strongly pushed forward as an archetype of heavy fermion driven by the local quantum criticality scenario \cite{Gegenwart2008, Si2013}. 
The proximity of YbRh$_2$Si$_2$ to AF quantum criticality is attested by the low value of its N\'eel temperature $T_N \sim 70$\,mK, far smaller than its Kondo temperature $T_K \sim 25$\,K \cite{Gegenwart2002}. Experimentally, a major appeal of this system is that  fine field tuning of the AF-PM border can be used to monitor the proximity to the QCP \cite{Gegenwart2008, Gegenwart2002}. Due to the large anisotropy of the magnetic properties, the critical field $H_c$ driving the AF-PM instability is also anisotropic, respectively 0.066\,T and 0.6\,T in the easy magnetization plane (\textbf{a},\textbf{b}), and along the \textbf{c} axis of the body centered tetragonal crystal. Evidence of FS reconstruction at $H_c$ are mainly based on Hall effect measurements \cite{Paschen2004, Friedemann2010a}. However, heavy fermions are multiband systems, so that the interpretation of the Hall effect is not straightforward. Other concomitant proof of the FS reconstruction in YbRh$_2$Si$_2$ must be found. No convincing changes of the FS at $H_c$ appear on the thermoelectric power \cite{Hartmann2010, Machida2012a}. 

Recently, the debate has been focused on the detection of a breakdown of the Fermi-liquid regime at $H_c$, using thermal transport at low temperature as a probe: A violation of the Wiedemann-Franz law (WFL) would point out a deep change of the electronic excitations responsible for charge and heat transport \cite{Pfau2012, Machida2013, Reid2014}. In metallic systems, at $T\to 0$~K, when quasiparticle scattering is dominated by elastic mechanisms
, all reported cases but one satisfy the WFL, stating that $L \to L_0$ in the low temperature limit, where $L=\kappa\rho/T$ is the Lorenz ratio and $L_0=L(T \to 0)$ is the Lorenz number ($L_0 = (\pi^2/3)(k_B/e)^2$). The only case reported so far that would not satisfy the WFL in this limit is CeCoIn$_5$, for the heat current along the \textbf{c}-axis \cite{Paglione2003}, which has been put forward as a demonstration of breakdown of the quasiparticle picture in a ``hot spot'' QCP scenario. But even in this last case, one cannot exclude that measurements below 30\,mK (the lowest temperature reached in \cite{Paglione2003}) would invalidate the violation of the WFL. 

Three different groups have recently published thermal conductivity ($\kappa$) experiments coupled with careful resistivity ($\rho$) experiments in YbRh$_2$Si$_2$, in order to study the temperature dependence of the Lorenz ratio $L(T) = \kappa \rho /T$. 
Contradictory conclusions have been drawn by the three groups. The first one \cite{Pfau2012} claims that the WFL is violated at $H_c$, as $L(T \to 0) \sim 0.90 L_0$, while the others concluded conservatively that $L(T \to 0)$ at $H_c$ is $L_0$ \cite{Machida2013,Reid2014}. However, all the experimental data of the three groups are in good agreement: The main differences lie in the interpretation, and in the lowest temperature reached, i.e. in the extrapolation of the zero temperature values. This last point is a very sensitive issue, as an additional contribution to the electronic thermal conductivity appears at very low temperatures (below 30\,mK) in YbRh$_2$Si$_2$, which is very difficult to separate from the quasiparticles contribution. The measurements we report in the present paper, were realized down to 10\,mK, twice lower than in those reported in \cite{Pfau2012}, and with a strong emphasis on the analysis of this extra contribution, notably to detect if anything specific happens at the critical field $H_c$.


The single crystal investigated in the present study was grown out of In flux. It is the same as used in Ref.~\cite{Pourret2013} for the thermoelectric power and magnetoresistance studies. The residual resistivity ratio RRR $=\rho (300{\rm K})/ \rho (0{\rm K})$ of the crystal is 65, the observation of quantum oscillations in the resistivity indicates its high quality. The magnetic field was applied parallel to the current (longitudinal configuration: \textbf{H}//\textbf{j}$_Q$//[1,1,0]), up to 4\,T. The zero field measurements have been performed with a zero field cooled magnet. At low field, a calibrated Hall sensor has been used to determine precisely the applied magnetic field. The thermal conductivity was measured using a standard two thermometers-one heater setup, and was checked to be independent of the heat gradient applied, with $\Delta T/T$ in the range 0.5-10\,\%. Calibration of the thermometers was realized in situ, against a CMN paramagnetic salt in zero field for temperatures below 0.1K and down to 7\,mK, and against reference thermometers (calibrated with the CMN) placed in the compensated zone of the magnet for field measurements. The electrical resistivity was measured simultaneously, with a conventional lock-in detection at low frequency (around 2\,Hz, for minimum quadrature signal), allowing a precise determination of $L(T)$.


\begin{figure}
		\includegraphics[width=0.75\textwidth]{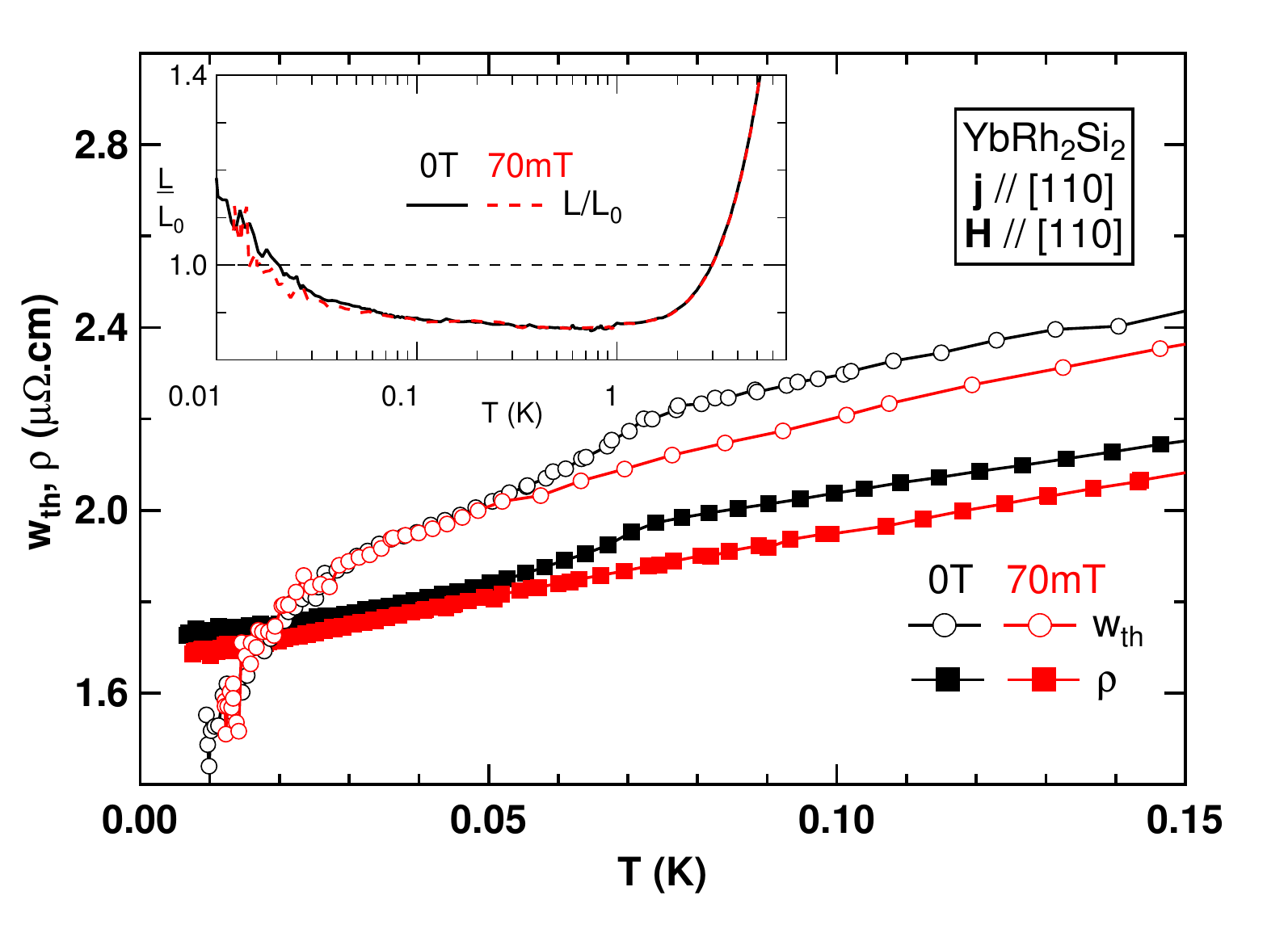}
	\caption{(color online) Electrical resistivity (full squares) and thermal resistivity (open circles) of YbRh$_2$Si$_2$ at 0\,T (in black) and at 70\,mT, close to $H_c$ (in red) at low temperature. Clear anomalies are observed in zero field at $T_N$, and a new channel for heat transport appears below 30\,mK (see text). \textit{Inset}: Normalized ratio $L(T)/L_0$ in the whole temperature range at 0\,T and 70\,mT.
	}
	\label{fig1}
\end{figure}

Figure \ref{fig1} compares the electrical resistivity $\rho$, and the thermal resistivity $w_{th}= \frac{L_0 T}{\kappa}$, at low temperature (below 0.15\,K), for zero magnetic field ($H$ = 0\,T) and very slightly above the critical field $H_c$ ($H$ = 70\,mT). This is the most important result, but let us first describe the insert which gives a broader view thanks to an extended temperature range (from 7\,mK up to 7\,K). It displays the Lorenz ratio $L(T)/L_0$ for both fields, with a logarithmic temperature scale: There is a large temperature range, from 3\,K down to 0.02\,K, where the normalized ratio $L(T)/L_0$ is clearly below 1, down to a value $L(T)/L_0 \approx 0.87$. This is indicative of dominant inelastic scattering (the so-called ``vertical processes''), which affect more strongly thermal than electrical transport. The rather unusual feature is that despite a large residual resistivity (the RRR is ``only'' 65 for our sample), inelastic scattering has a significant contribution down to such low temperatures. But this is consistent with the marked temperature dependence of the resistivity down to the lowest measured temperatures (8\,mK), as can be seen on the main graph. Inversely, the fast increase of the Lorenz ratio above 2\,K, overpassing 1 above 3\,K, indicates additional contributions to the heat transport (like phonons). 

These ``bosonic'' contributions are expected to vanish rapidly below 1\,K, and therefore, the new increase of the Lorenz ratio above 1 below 20\,mK is again a very unusual feature, which could not be anticipated from the resistivity behavior. This low-temperature increase of the Lorenz ratio has been already observed below 30\,mK by Pfau \textit{et al.} \cite{Pfau2012}, on a slightly more resistive sample (RRR $\approx 45$ instead of 65). But we insist that this is probably the most puzzling feature of thermal transport in YbRh$_2$Si$_2$: How is it that new bosonic excitations, able to contribute as a new heat channel, appear at such low temperatures? Moreover, contrary to the statement in \cite{Pfau2012}, we observe that this increase is essentially unchanged for a field of 70\,mT, suppressing the long range AF order, and the putative magnon (long lived) excitations expected in this ordered phase. Discussion of the origin and field evolution of this contribution is continued later in this paper.


The main plot of Fig. \ref{fig1} shows the detailed evolution of $\rho$ and $w_{th}$ below 0.15\,K, down to 10\,mK: In particular, one clearly observes a kink at $T_N$ in zero field on the electrical and thermal transport. It is suppressed by an applied field of 70\,mT (very close to $H_c$), which also lowers both resistivities. It demonstrates that the critical fluctuations emerging in the neighborhood of the AF transition contribute to the scattering of the quasiparticles. However, this scattering seems to be rapidly damped below $T_N$, as no change of $w_{th}$ is observed below 50\,mK, between zero field and 70\,mT. Concerning now the additional low temperature contribution, the strong decrease of $w_{th}$ below 30\,mK appears independent of the applied field. It already shows the difficulty to discuss the validity of the WFL: The linear extrapolation of the data of $w_{th}$ between 30 and 50\,mK is compatible, within the dispersion of the points, with that of the resistivity at 0\,K. The whole question is to control these extrapolations, or to estimate the amplitude and range of the extra contribution responsible for the low temperature drop of $w_{th}$. But in any case, on the bare data, no drastic change is seen between zero field and $H_c$.

Figure \ref{fig3} compares the relative temperature variation of $L(T)/L_0$ for fields above $H_c$, up to 4\,T (see Supplementary Material A, figure \ref{figS2} for the data of $\kappa$ and $\rho$). It appears that the low temperature increase of the Lorenz ratio is field sensitive above 100\,mT, and rapidly suppressed for rather weak fields (above 0.2\,T). This supports a magnetic origin for this additional heat channel, which is reasonable as we did not expect a sizable phonon contribution at such low temperatures. In parallel, we do observe that the decrease below one of the Lorenz ratio above 30\,mK is also progressively suppressed under field, pointing to a magnetic origin of the low temperature inelastic scattering. However under field, for H $>0.5$\,T (see also insert up to 4\,T), $L(T\to0)$ seems to saturate at 0.97 $L_0$. 
This small deviation from $L_0$ may be due to experimental precision, or related to the complex multiband structure of the FS. In any case, the fact that this value does not change with fields much larger than $H_c$ means that it cannot be seen as a violation of the WFL at the QCP. In the following, particularly to estimate the electronic contribution, the Lorenz number is taken at this value found for $T \to$ 0\,K: $L(0)=0.97L_0$. 

\begin{figure}
	\centering
		\includegraphics[width=0.75\textwidth]{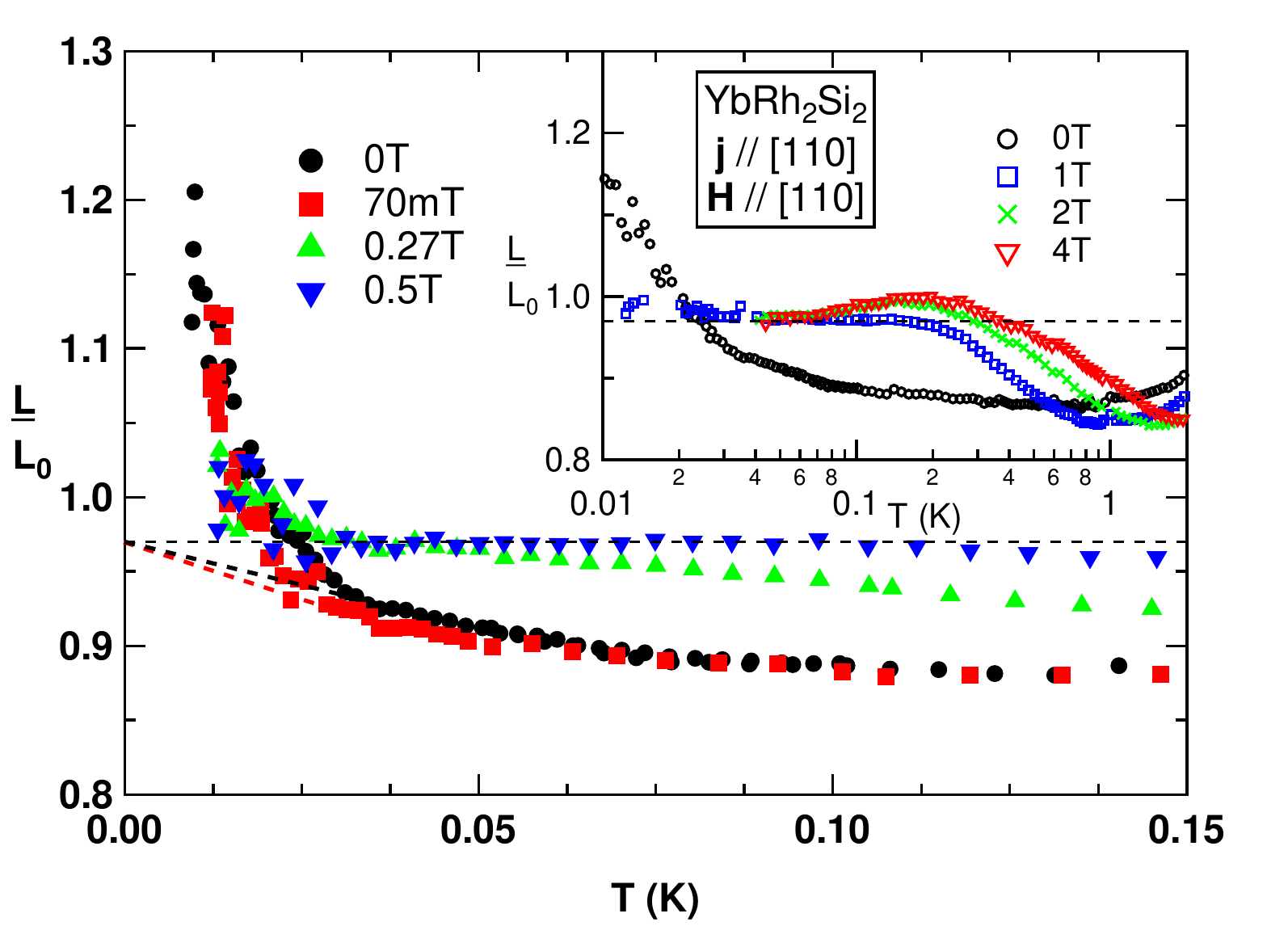}
	\caption{(color online) Ratio $L(T)/L_0$ at low temperatures below 1\,T. The dashed lines are the extrapolation to $L(0)=0.97L_0$ when T $>$ 30\,mK. In inset, $L(T)/L_0$ at higher fields, which extrapolates also to 0.97 at low temperature.}
	\label{fig3}
\end{figure}

\begin{figure*}
	\centering
		\includegraphics[width=0.95\textwidth]{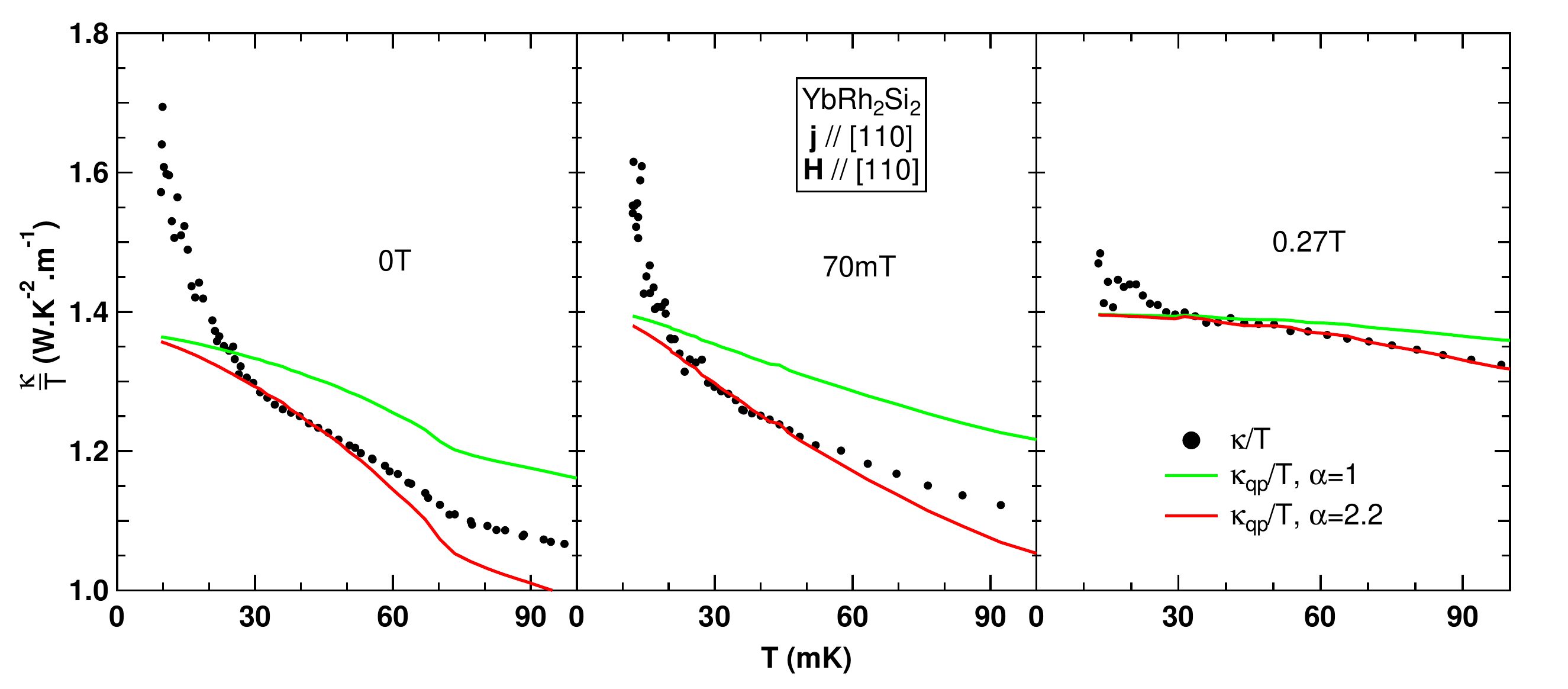}
	\caption{(color online) Comparison of the electronic thermal conductivity using the expression \ref{eq1} with $\alpha=1$ and $\alpha=2.2$. The latter value was chosen to optimize the low temperature contribution of $\kappa_{qp}$.}
	\label{fig4}
\end{figure*}

As regards the question of the origin of the additional contribution to thermal transport appearing below 30\,mK (hereafter labelled $\kappa_{add}$), all hints point to a magnetic origin: (i) even if it survives the critical field, it is rapidly suppressed by fields of order of 0.5\,T; (ii) a new phononic contribution is completely unexpected in this temperature range: It would require a strong lattice instability, which has not been detected so far; (iii) besides the antiferromagnetic long range order at 70\,mK, other magnetic transitions have been reported at 12\,mK \cite{Schuberth2009} and/or 2\,mK \cite{Schuberth2013}. In reference \cite{Pfau2012}, $\kappa_{add}$ has been attributed to magnons, an assumption supported by comparison to the specific heat ($C_p$) data: The temperature dependence of $C_p$ has a bare $T^3$ regime below $T^\star = 55$\,mK \cite{Gegenwart2008, Gegenwart2002}, which is attributed to AF magnons. Because it exists even when the long range ordered state is suppressed, we would rather compare $\kappa_{add}$ in YbRh$_2$Si$_2$ to a similar additional contribution to heat transport observed in the ferromagnet UCoGe, both above and below T$_{Curie}$, and also in a very low temperature range (for the heat current along the \textbf{b}-axis) \cite{Taupin2014}. Obviously, ``paramagnons'' or overdamped spin fluctuations, can also contribute to an additional heat transport channel.

Quantitatively, we will perform a similar analysis as in \cite{Taupin2014}, assuming that we can separate the electronic contribution in a quasiparticle $\kappa_{qp}$ channel, which should follow (or not!) the WFL, and a ``spin-fluctuation'' contribution $\kappa_{add}$, expected for example to be much less sensitive to the RRR. As for the case of UCoGe, a difficulty for quantitative estimation of $\kappa_{add}$ is that the quasiparticle contribution is badly known, due to the importance of inelastic scattering even at 30\,mK (see the Lorenz ratio in figure \ref{fig1}). So we assume that $\kappa_{qp}$ can be described by \cite{Wagner1971}:
\begin{equation}
	\frac{\kappa_{qp}}{T} = \frac{{L(0)}}{\rho_0 + \alpha (\rho -\rho_0)} 
	\label{eq1}
\end{equation}

This expression takes into account the stronger effect of the electronic inelastic scattering on the thermal transport (than on the electrical transport), through the factor $\alpha$. $\rho_0$ is the residual resistivity and $\alpha=1+W_{vert}/W_{hor}$, where $W_{vert}$ and $W_{hor}$ are respectively the scattering rates due to vertical (small wave-vector transfer $q$, inelastic) and horizontal (large $q$, mostly elastic) processes, assuming that Mathiessen's rule holds. The crude simplification is to take $\alpha$ constant, which can be valid for temperatures much smaller than the typical energy of the fluctuations responsible for the inelastic scattering. This simplification has already been widely used for the heavy fermion systems like UPt$_3$ \cite{Lussier1994}, CeRhIn$_5$ \cite{Paglione2005} and UCoGe \cite{Taupin2014}. The figure \ref{fig4} compares the bare data with different estimates of $\kappa_{qp}$ for $\alpha = 1$ i.e $L(T) = L(0)$ and $\alpha = 2.2$, at 0\,T (in the AF state), 70\,mT (close to $H_c$) and 0.27\,T (in the PM state). Clearly, $\alpha = 1$ overestimates the quasiparticle contribution, whereas $\alpha = 2.2$ (the best value we found) gives an estimate of $\kappa_{qp}$ compatible with the data for the three fields. It even gives a fair account of the data between 30\,mK and 100\,mK at 0.27\,T. The extracted extra contribution is shown in the Supplementary Material A, figure \ref{figS3}.

The most important information from the analysis shown on figure \ref{fig4} is that little change is observed between zero field and $H_c$, apart from the suppression of the anomaly at $T_N$. Otherwise, (i) it is possible to analyze the data of thermal transport in YbRh$_2$Si$_2$, assuming that the quasiparticle contribution follows the WFL even at $H_c$, and with the same coefficient $\alpha=2.2$ describing the effect of inelastic scattering; (ii) $\kappa_{add}$ is little influenced by the suppression of the N\'eel temperature at $H_c$, and really starts to decrease only above this field (see also figure \ref{figS3} of the Supplementary Material A), pointing to a contribution of driven (robust) short range correlations.

For point (i), it will be very difficult to reach firmer conclusions, as even lower temperature measurements would face the problem of the dominant additional contributions, and of the difficulty of a robust quantitative analysis, together with the regimes introduced by the additional transitions at 12 and 2\,mK \cite{Schuberth2009, Schuberth2013}. For point (ii), this is similar to observations on the thermopower, where the change of regime under field occurs far above $H_c$ \cite{Hartmann2010, Machida2012a}, and little is seen when crossing $H_c$. 

Moreover, the question of the role of magnetic fluctuations in the heat transport in YbRh$_2$Si$_2$ is not limited to the low temperature regime. Also at high temperatures, above 1K, we can detect a field dependence of $w_{th}$, which (like for resistivity), points to a contribution of magnetic fluctuations on the scattering mechanisms. In this temperature range, microscopic measurements have revealed a complex situation: NMR experiments \cite{Ishida2002} have shown that AF fluctuations with finite wavevectors compete with ferromagnetic (FM) fluctuations; recent investigations by NMR \cite{Kambe2014a} gives the image of interfering Fermi liquid and non Fermi liquid component close to $H_c$. Neutron scattering experiments \cite{Stock2012} down to 0.1\,K up to 10\,T along the \textbf{c}-axis confirm that incommensurate AF correlations become dominant on cooling against the FM correlations, while the magnetic field induces a Zeeman resonant excitation. These FM fluctuations could be an important ingredient to understand the observation of a well defined electron spin resonance in YbRh$_2$Si$_2$ \cite{Sichelschmidt2003,Abrahams2008}. So at least far from $T_N$ or at high magnetic field, FM coupling may be dominant, pointing to the proximity of a FM QCP \cite{Knebel2006a, Lausberg2013}. We do find (see Supplementary material B) that the thermal resistivity above 1K behaves as predicted by Ueda and Moriya \cite{Ueda1975} when inelastic scattering is controlled by ferromagnetic fluctuations.  


To summarize we have investigated in detail the thermal transport of YbRh$_2$Si$_2$, in the AF phase and in the PM phase, below and above $H_c$, and extended the measurements down to 10\,mK. This study clearly shows that an additional magnetic contribution to thermal conductivity appears below 30\,mK, with a gradual decrease for fields above the critical field $H_c$. This has been deduced with an electronic quasiparticle contribution, which also has no peculiar singularity at $H_c$, and satisfies the WFL even at $H_c$. To conclude unambiguously, by direct measurements, if the WFL is satisfied or violated, would require measurements down to the mK temperature range, where the additional contribution should eventually disappear. But other recent experiments also challenge the ``local QCP scenario'', which could have led to the violation of the WFL at the critical field: For example, ARPES measurements \cite{Guttler2014,Danzenbacher2011} show that above $T_N$, the 4f-electrons in the paramagnetic phase are clearly itinerant, which leaves open the validity of the image of small and large Fermi surface fluctuations developed in the Kondo breakdown scenario. A new appealing approach is the so-called ``strong coupling theory'' of heavy-fermion quantum criticality, where large critical spin fluctuations at the AF wave-vector induce fluctuations at small wave-vectors, producing a diverging effective mass over the entire Fermi surface \cite{Abrahams2014}. It would be interesting to see if these fluctuations could be responsible for the detected low temperature contribution, as well as to the large inelastic scattering observed down to very low temperatures ?



This work has been supported by the French ANR grant SINUS and the ERC grant ``NewHeavyFermion''.



\input{YbRh2Si2.bbl}

\newpage
\part{Supplementary Materials}
\section{Supplementary Material A: Details on the transport measurements}
\begin{figure}[ht!]
	\centering
		\includegraphics[width=0.75\textwidth]{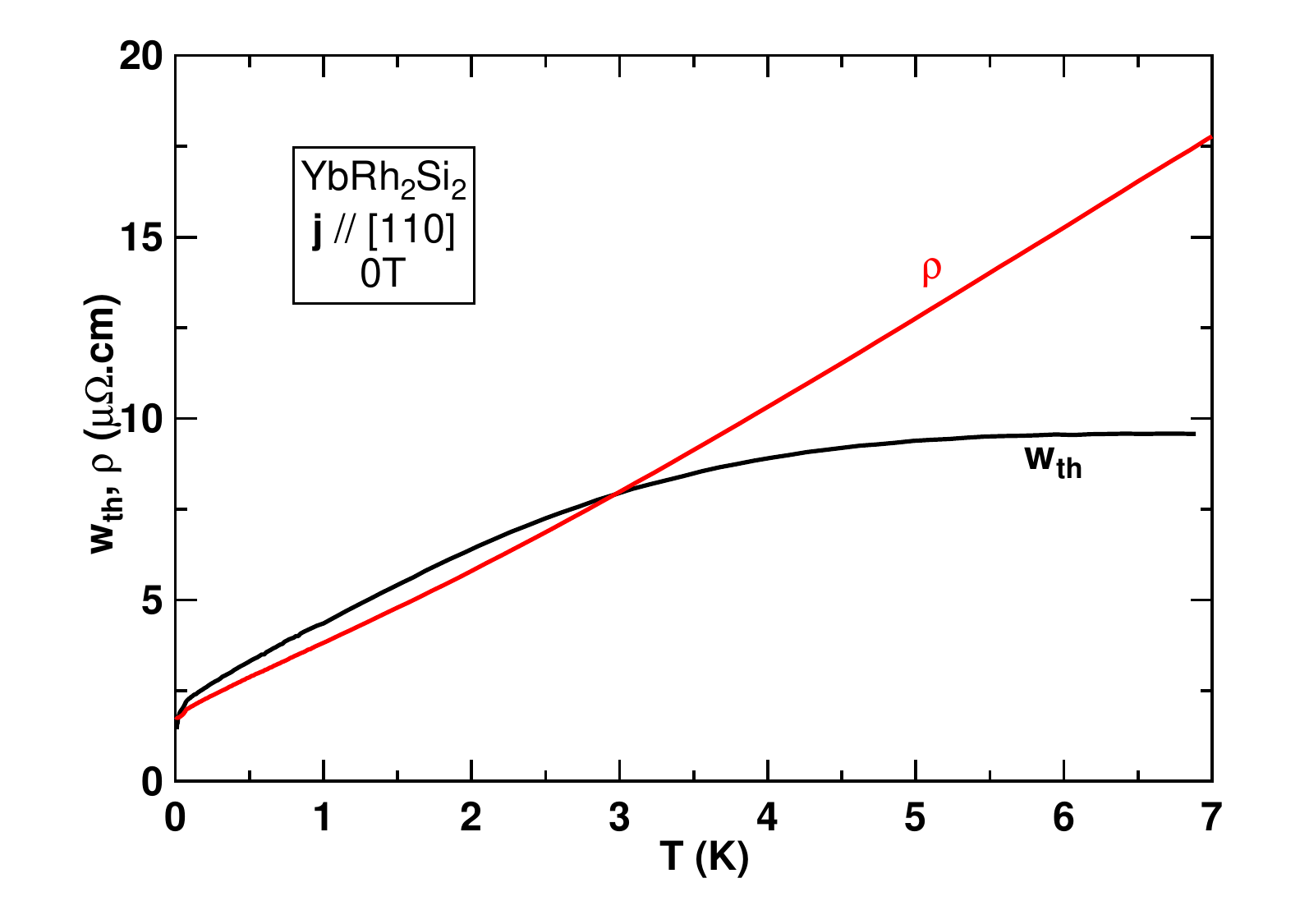}
	\caption{(color online) Electrical and thermal resistivities at 0\,T in the whole temperature range from 10\,mK up to 7\,K.}
	\label{figS1}
\end{figure}

As a complement to figure \ref{fig1} of the main paper, displaying the thermal and electrical resistivities below 0.15\,K, figure \ref{figS1} presents $\rho$ and $w_{th}$ at 0\,T up to 7\,K: This shows the strong temperature dependence of $\rho$ in the whole temperature range, as well as the saturation of $w_{th}$ above 3\,K, due to the growing contribution of bosonic excitations to heat transport.

\begin{figure}[ht!]
	\centering
		\includegraphics[width=0.75\textwidth]{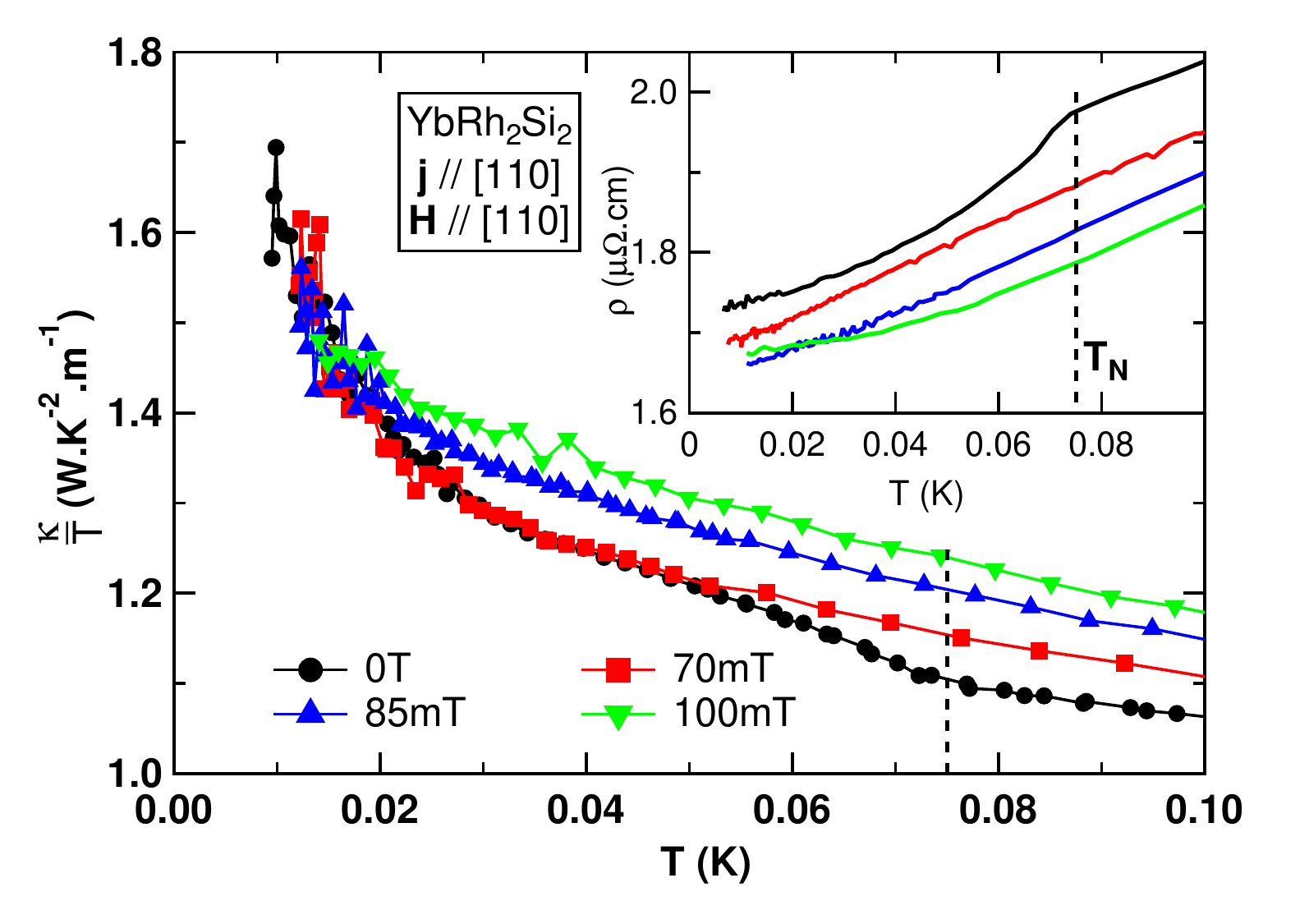}
	\caption{(color online) Thermal conductivity of YbRh$_2$Si$_2$ at low temperature below 0.1\,T. In inset, the corresponding electrical resistivity. The dash line shows the N\'eel temperature at zero field.}
	\label{figS2}
\end{figure}

If several (parallel) channels contribute to the thermal transport, it is better to discuss the thermal conductivity rather than $w_{th}$. It is displayed on figure \ref{figS2}, at low temperature and up to 0.1\,T, and the corresponding electrical resistivity is shown in the inset. The smooth variation of the thermal conductivity through $H_c$ is confirmed by the temperature dependence of $\kappa/T$ below 100\,mK on each side of $H_c$ for different magnetic fields.

\begin{figure}[ht!]
	\centering
		\includegraphics[width=0.75\textwidth]{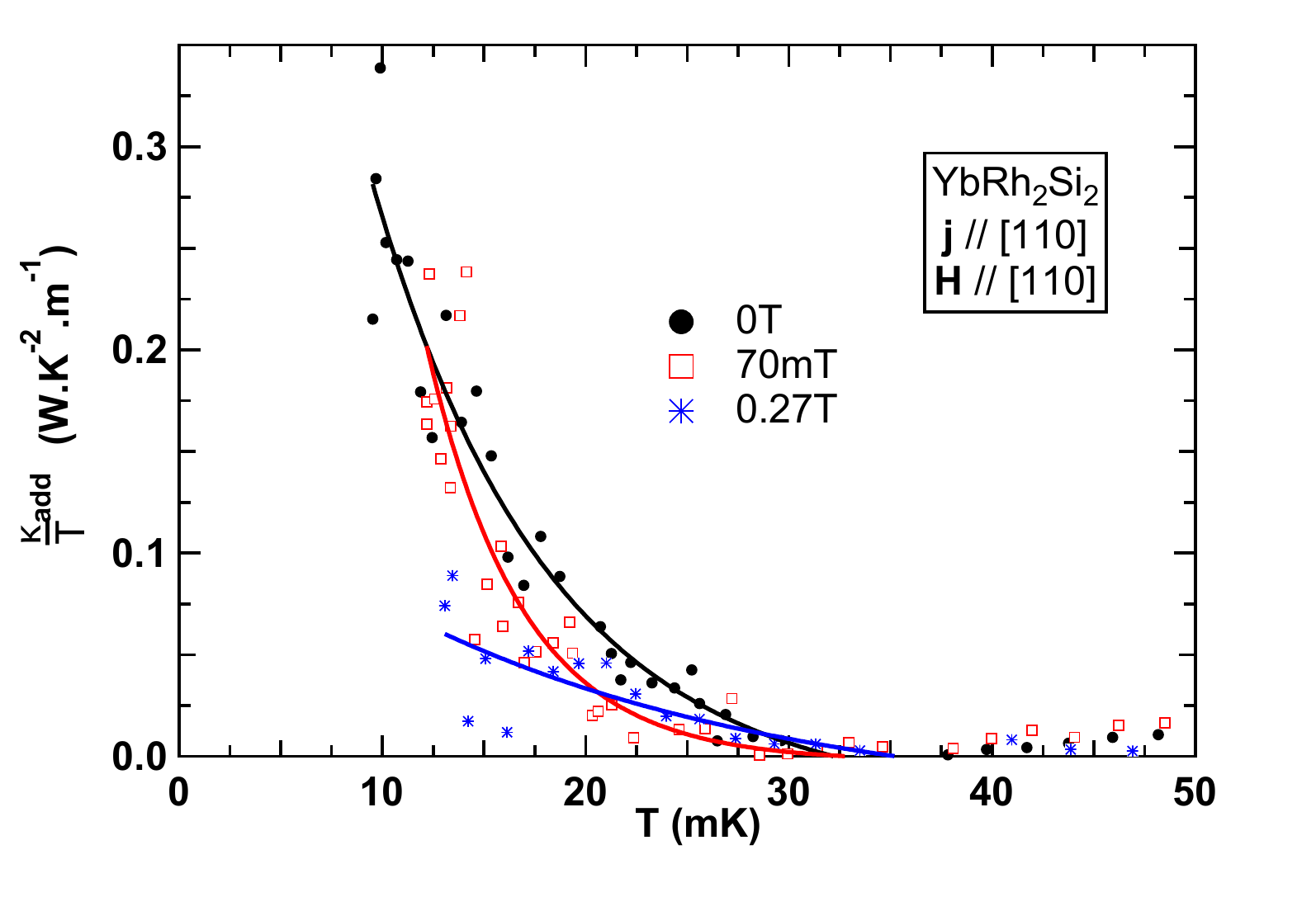}
	\caption{(color online) Estimation of the very low temperature extra contribution to the thermal conductivity at 0\,T, 70\,mT and 0.27\,T. No significant changes are seen between 0\,T and 70\,mT, indicating that $\kappa_{add}$ has certainly the same origin at both fields. $\kappa_{add}$ exists for fields higher than $H_c$, but with a decreasing amplitude. Lines are guides to the eyes.}
	\label{figS3}
\end{figure}

As explained in the main paper, we could analyze the quasiparticle contribution, taking into account the peculiar effects of inelastic scattering on heat transport, and with no detected anomaly when crossing the critical field $H_c$. Figure \ref{figS3} presents the additional contribution $\kappa_{add}/T$ for three different fields H = 0, H $ \sim H_c$ and H $\sim 4 H_c$, subtracting this calculated value of $\kappa_{qp}(T)$. As announced, there is little change of $\kappa_{add}$ at $H_c$, whereas its collapse is noticeable for a field of 0.27\,T. For even higher field values, $\kappa_{add}$ rapidly falls bellow the error bars of the analysis, an effect reinforced by the increase of the bottom temperature of the measurements due to prohibitive equilibrium relaxation times. But as regards the direct test of the validity of the WFL, the strong ``divergence'' of  $\kappa_{add}/T$ at low field at 10\,mK sets the level of the experimental challenge: Probably temperatures of order a few mK at most, are required to observe the downturn of this contribution?
 
\section{Supplementary Material B: Details on the high temperature analysis}
As underlined, evidences exist that the main magnetic intersite interaction is temperature dependent. As the ``average'' effective mass increases drastically on cooling, the relative weight of the different outbands may change and produce a feedback on the magnetic coupling itself. Due to evidences for FM fluctuations above 1\,K \cite{Kambe2014a,Stock2012}, we investigated if YbRh$_2$Si$_2$ could be seen, far above $T_N$, as a weak ferromagnet above its Curie temperature from the point of view of thermal transport? In this case, if only the quasiparticle-magnetic fluctuations interactions are taken into account, Ueda and Moriya have predicted that the 
 thermal resistance should behave as \cite{Ueda1975}:
\begin{eqnarray}
		w_{th} &=& \rho+BT
	\label{eqS1}
\end{eqnarray}

So, even if the electrical resistivity has no real simple temperature dependence (see figures 3-4 of reference \cite{Ueda1975}, the additional thermal resistance for weak ferromagnets above $T_{Curie}$ or near ferromagnets at high enough temperatures is roughly linear in temperature (with our definition of $w_{th}$): See figure 5 of \cite{Ueda1975}. Naturally, in this temperature range, we also expect a phononic contribution, limited by electron-phonon scattering, which should follow a quadratic temperature dependence: $\kappa_{ph}=PT^2$, with $P$ temperature and field independent. We obtain thus for the thermal conductivity at high temperature:
\begin{eqnarray}
	\kappa &=& \kappa_{qp}+\kappa_{ph}\\
		&=& \frac{L_0 T}{\rho + BT} + PT^2
	\label{eqS2}
\end{eqnarray}

Once the phonons are subtracted, the thermal resistance $w_{th}\equiv w_{qp}=\frac{L_0 T}{\kappa_{qp}}$ represents the electronic heat channel with the scattering of quasiparticles by FM fluctuations. We call the additional linear term in the thermal resistivity $w_{qp-fluc} = BT$. Contrary to the low temperature situation, the magnetic fluctuations above 1\,K do not contribute with their own channel to heat transport: They only have the ``usual'' contribution to the inelastic scattering of the quasiparticles. The expression \ref{eqS2} is very simple, as $\rho$ is determined experimentally and the parameter $P$ is field independent. Only the parameter $B$ can vary with the field. Such a decomposition of the thermal conductivity has already been successfully used for the weak ferromagnet ZrZn$_2$ \cite{Smith2008}. 

\begin{figure}
	\centering
		\includegraphics[width=0.75\textwidth]{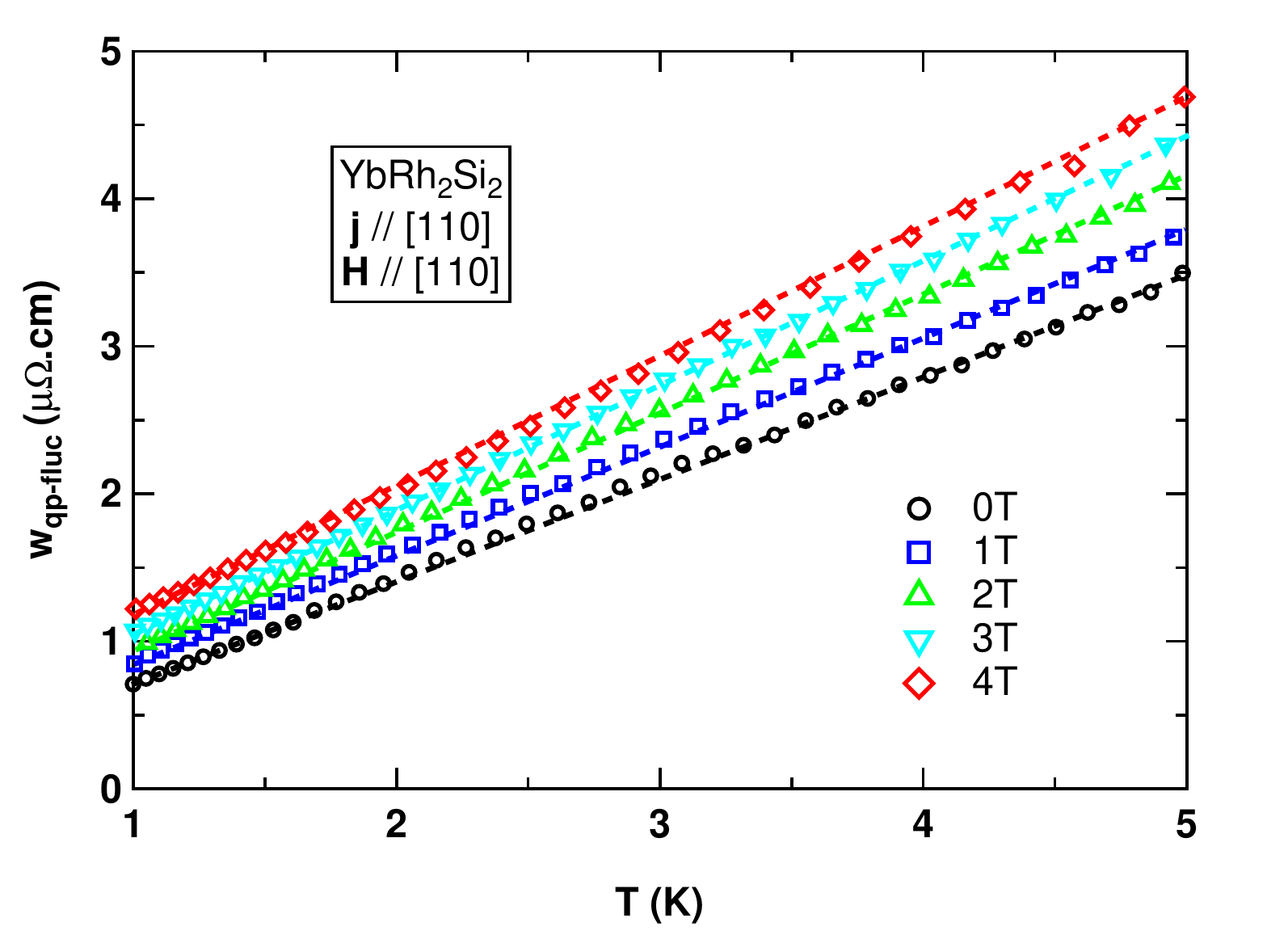}
	\caption{(color online) Electronic thermal resistivity $w_{qp-fluc}$ up to 5\,K after subtraction of the electrical resistivity. For clarity, an offset has been added to each curve ($+0.2$\,$\mu \Omega$.cm at each field step). Dashed lines are guides to the eyes.}
	\label{figS4}
\end{figure}

Surprisingly, after removal of the phonon and resistivity contribution, we do find that $w_{qp-fluc}$ is linear above 1\,K, for fields up to 4\,T, as shown in figure \ref{figS4}, for a field independent parameter $P=0.022$\,W.K$^{-3}$.m$^{-1}$, close to 0.017\,W.K$^{-3}$.m$^{-1}$ used by Pfau \textit{et al.} \cite{Pfau2012}. At lower temperatures, a deviation of $w_{qp-fluc}$ from linearity appears, which may be due to the growing importance of AF fluctuations.

\begin{figure}[ht!]
	\centering
		\includegraphics[width=0.75\textwidth]{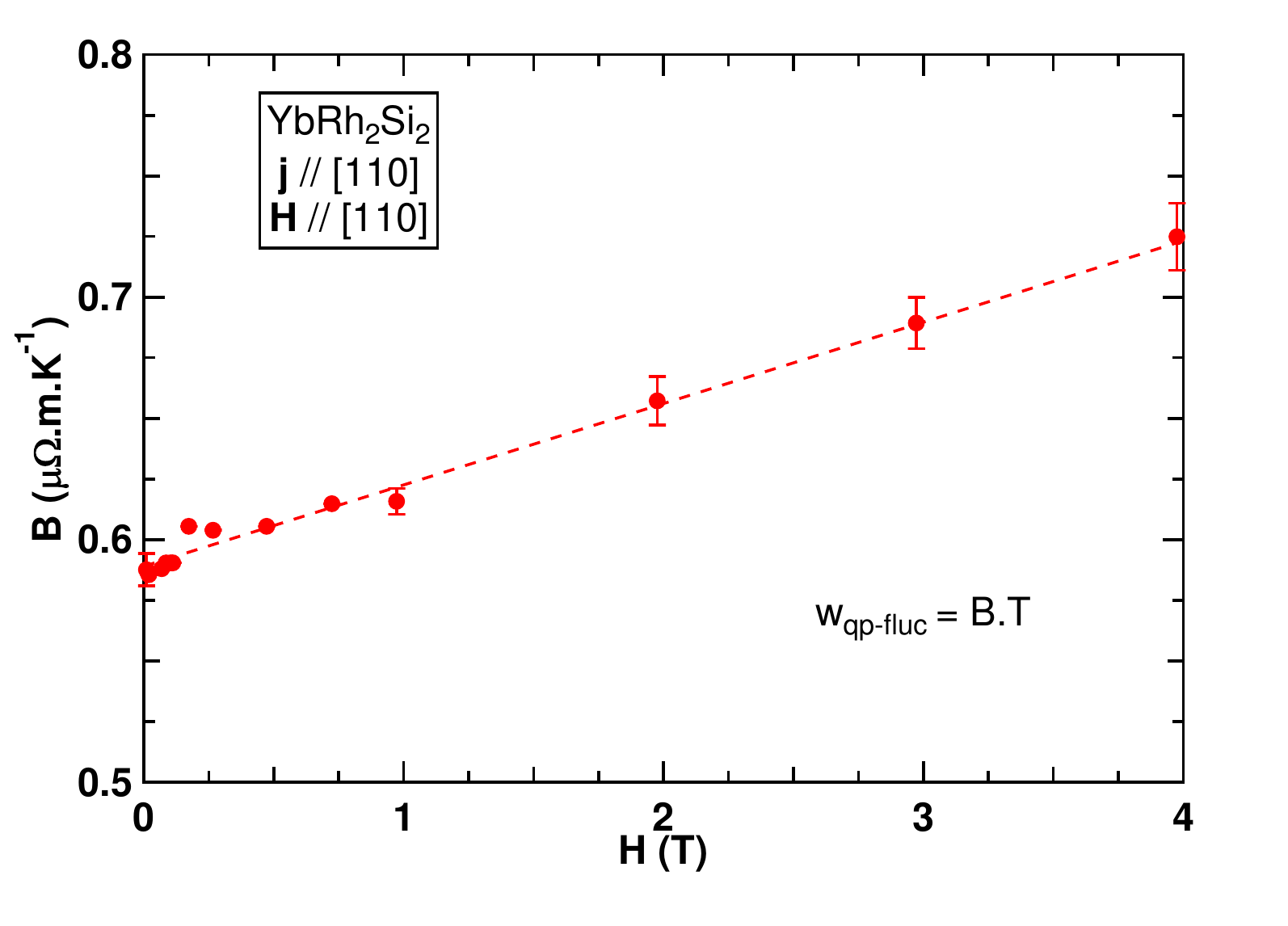}
	\caption{(color online) Evolution of the coefficient $B$ with the magnetic field.}
	\label{figS5}
\end{figure}

The evolution of the $B$ coefficient is shown in figure \ref{figS5}: It increases linearly with the magnetic field, as if the applied field and the warming reinforce the ferromagnetic fluctuations. This behaviour is quite singular and may be related to the intermediate valent character of YbRh$_2$Si$_2$. In Ce compounds, like in the CeRu$_2$Si$_2$ family, ferromagnetic fluctuations are just enhanced at the metamagnetic field where simultaneously the AF interactions drops \cite{Flouquet2004}.




\input{YbRh2Si2_SM.bbl}

\end{document}

%% file: YbRh2Si2.bbl
%

%% file: YbRh2Si2_SM.bbl
%

%% file: YbRh2Si2_vArXiv.bbl
\begin{thebibliography}{31}%
\makeatletter
\providecommand \@ifxundefined [1]{%
 \@ifx{#1\undefined}
}%
\providecommand \@ifnum [1]{%
 \ifnum #1\expandafter \@firstoftwo
 \else \expandafter \@secondoftwo
 \fi
}%
\providecommand \@ifx [1]{%
 \ifx #1\expandafter \@firstoftwo
 \else \expandafter \@secondoftwo
 \fi
}%
\providecommand \natexlab [1]{#1}%
\providecommand \enquote  [1]{``#1''}%
\providecommand \bibnamefont  [1]{#1}%
\providecommand \bibfnamefont [1]{#1}%
\providecommand \citenamefont [1]{#1}%
\providecommand \href@noop [0]{\@secondoftwo}%
\providecommand \href [0]{\begingroup \@sanitize@url \@href}%
\providecommand \@href[1]{\@@startlink{#1}\@@href}%
\providecommand \@@href[1]{\endgroup#1\@@endlink}%
\providecommand \@sanitize@url [0]{\catcode `\\12\catcode `\$12\catcode
  `\&12\catcode `\#12\catcode `\^12\catcode `\_12\catcode `\%12\relax}%
\providecommand \@@startlink[1]{}%
\providecommand \@@endlink[0]{}%
\providecommand \url  [0]{\begingroup\@sanitize@url \@url }%
\providecommand \@url [1]{\endgroup\@href {#1}{\urlprefix }}%
\providecommand \urlprefix  [0]{URL }%
\providecommand \Eprint [0]{\href }%
\providecommand \doibase [0]{http://dx.doi.org/}%
\providecommand \selectlanguage [0]{\@gobble}%
\providecommand \bibinfo  [0]{\@secondoftwo}%
\providecommand \bibfield  [0]{\@secondoftwo}%
\providecommand \translation [1]{[#1]}%
\providecommand \BibitemOpen [0]{}%
\providecommand \bibitemStop [0]{}%
\providecommand \bibitemNoStop [0]{.\EOS\space}%
\providecommand \EOS [0]{\spacefactor3000\relax}%
\providecommand \BibitemShut  [1]{\csname bibitem#1\endcsname}%
\let\auto@bib@innerbib\@empty
\bibitem [{\citenamefont {L\"ohneysen}\ \emph {et~al.}(2007)\citenamefont
  {L\"ohneysen}, \citenamefont {Rosch}, \citenamefont {Vojta},\ and\
  \citenamefont {W\"olfle}}]{Lohneysen2007}%
  \BibitemOpen
  \bibfield  {author} {\bibinfo {author} {\bibfnamefont {H.~v.}\ \bibnamefont
  {L\"ohneysen}}, \bibinfo {author} {\bibfnamefont {A.}~\bibnamefont {Rosch}},
  \bibinfo {author} {\bibfnamefont {M.}~\bibnamefont {Vojta}}, \ and\ \bibinfo
  {author} {\bibfnamefont {P.}~\bibnamefont {W\"olfle}},\ }\href {\doibase
  10.1103/RevModPhys.79.1015} {\bibfield  {journal} {\bibinfo  {journal} {Rev.
  Mod. Phys.}\ }\textbf {\bibinfo {volume} {79}},\ \bibinfo {eid} {1015}
  (\bibinfo {year} {2007})}\BibitemShut {NoStop}%
\bibitem [{\citenamefont {Coleman}\ \emph {et~al.}(2001)\citenamefont
  {Coleman}, \citenamefont {P\'{e}pin}, \citenamefont {Si},\ and\ \citenamefont
  {Ramazashvili}}]{Coleman2001}%
  \BibitemOpen
  \bibfield  {author} {\bibinfo {author} {\bibfnamefont {P.}~\bibnamefont
  {Coleman}}, \bibinfo {author} {\bibfnamefont {C.}~\bibnamefont {P\'{e}pin}},
  \bibinfo {author} {\bibfnamefont {Q.}~\bibnamefont {Si}}, \ and\ \bibinfo
  {author} {\bibfnamefont {R.}~\bibnamefont {Ramazashvili}},\ }\href
  {http://iopscience.iop.org/0953-8984/13/35/202} {\bibfield  {journal}
  {\bibinfo  {journal} {J. Phys.: Condens. Matter}\ }\textbf {\bibinfo {volume}
  {13}},\ \bibinfo {pages} {R723} (\bibinfo {year} {2001})}\BibitemShut
  {NoStop}%
\bibitem [{\citenamefont {Gegenwart}\ \emph {et~al.}(2008)\citenamefont
  {Gegenwart}, \citenamefont {Si},\ and\ \citenamefont
  {Steglich}}]{Gegenwart2008}%
  \BibitemOpen
  \bibfield  {author} {\bibinfo {author} {\bibfnamefont {P.}~\bibnamefont
  {Gegenwart}}, \bibinfo {author} {\bibfnamefont {Q.}~\bibnamefont {Si}}, \
  and\ \bibinfo {author} {\bibfnamefont {F.}~\bibnamefont {Steglich}},\ }\href
  {\doibase 10.1038/nphys892} {\bibfield  {journal} {\bibinfo  {journal}
  {Nature Physics}\ }\textbf {\bibinfo {volume} {4}},\ \bibinfo {pages} {186}
  (\bibinfo {year} {2008})}\BibitemShut {NoStop}%
\bibitem [{\citenamefont {Si}\ and\ \citenamefont {Paschen}(2013)}]{Si2013}%
  \BibitemOpen
  \bibfield  {author} {\bibinfo {author} {\bibfnamefont {Q.}~\bibnamefont
  {Si}}\ and\ \bibinfo {author} {\bibfnamefont {S.}~\bibnamefont {Paschen}},\
  }\href {\doibase 10.1002/pssb.201300005} {\bibfield  {journal} {\bibinfo
  {journal} {Physica Status Solidi (b)}\ }\textbf {\bibinfo {volume} {250}},\
  \bibinfo {pages} {425} (\bibinfo {year} {2013})}\BibitemShut {NoStop}%
\bibitem [{\citenamefont {Gegenwart}\ \emph {et~al.}(2002)\citenamefont
  {Gegenwart}, \citenamefont {Custers}, \citenamefont {Geibel}, \citenamefont
  {Neumaier}, \citenamefont {Tayama}, \citenamefont {Tenya}, \citenamefont
  {Trovarelli},\ and\ \citenamefont {Steglich}}]{Gegenwart2002}%
  \BibitemOpen
  \bibfield  {author} {\bibinfo {author} {\bibfnamefont {P.}~\bibnamefont
  {Gegenwart}}, \bibinfo {author} {\bibfnamefont {J.}~\bibnamefont {Custers}},
  \bibinfo {author} {\bibfnamefont {C.}~\bibnamefont {Geibel}}, \bibinfo
  {author} {\bibfnamefont {K.}~\bibnamefont {Neumaier}}, \bibinfo {author}
  {\bibfnamefont {T.}~\bibnamefont {Tayama}}, \bibinfo {author} {\bibfnamefont
  {K.}~\bibnamefont {Tenya}}, \bibinfo {author} {\bibfnamefont
  {O.}~\bibnamefont {Trovarelli}}, \ and\ \bibinfo {author} {\bibfnamefont
  {F.}~\bibnamefont {Steglich}},\ }\href {\doibase
  10.1103/PhysRevLett.89.056402} {\bibfield  {journal} {\bibinfo  {journal}
  {Phys. Rev. Lett.}\ }\textbf {\bibinfo {volume} {89}},\ \bibinfo {pages}
  {056402} (\bibinfo {year} {2002})}\BibitemShut {NoStop}%
\bibitem [{\citenamefont {Paschen}\ \emph {et~al.}(2004)\citenamefont
  {Paschen}, \citenamefont {Luhmann}, \citenamefont {Wirth}, \citenamefont
  {Gegenwart}, \citenamefont {Trovarelli}, \citenamefont {Geibel},
  \citenamefont {Steglich}, \citenamefont {Coleman},\ and\ \citenamefont
  {Si}}]{Paschen2004}%
  \BibitemOpen
  \bibfield  {author} {\bibinfo {author} {\bibfnamefont {S.}~\bibnamefont
  {Paschen}}, \bibinfo {author} {\bibfnamefont {T.}~\bibnamefont {Luhmann}},
  \bibinfo {author} {\bibfnamefont {S.}~\bibnamefont {Wirth}}, \bibinfo
  {author} {\bibfnamefont {P.}~\bibnamefont {Gegenwart}}, \bibinfo {author}
  {\bibfnamefont {O.}~\bibnamefont {Trovarelli}}, \bibinfo {author}
  {\bibfnamefont {C.}~\bibnamefont {Geibel}}, \bibinfo {author} {\bibfnamefont
  {F.}~\bibnamefont {Steglich}}, \bibinfo {author} {\bibfnamefont
  {P.}~\bibnamefont {Coleman}}, \ and\ \bibinfo {author} {\bibfnamefont
  {Q.}~\bibnamefont {Si}},\ }\href {http://dx.doi.org/10.1038/nature03129}
  {\bibfield  {journal} {\bibinfo  {journal} {Nature}\ }\textbf {\bibinfo
  {volume} {432}},\ \bibinfo {pages} {881} (\bibinfo {year}
  {2004})}\BibitemShut {NoStop}%
\bibitem [{\citenamefont {Friedemann}\ \emph {et~al.}(2010)\citenamefont
  {Friedemann}, \citenamefont {Oeschler}, \citenamefont {Wirth}, \citenamefont
  {Krellner}, \citenamefont {Geibel}, \citenamefont {Steglich}, \citenamefont
  {Paschen}, \citenamefont {Kirchner},\ and\ \citenamefont
  {Si}}]{Friedemann2010a}%
  \BibitemOpen
  \bibfield  {author} {\bibinfo {author} {\bibfnamefont {S.}~\bibnamefont
  {Friedemann}}, \bibinfo {author} {\bibfnamefont {N.}~\bibnamefont
  {Oeschler}}, \bibinfo {author} {\bibfnamefont {S.}~\bibnamefont {Wirth}},
  \bibinfo {author} {\bibfnamefont {C.}~\bibnamefont {Krellner}}, \bibinfo
  {author} {\bibfnamefont {C.}~\bibnamefont {Geibel}}, \bibinfo {author}
  {\bibfnamefont {F.}~\bibnamefont {Steglich}}, \bibinfo {author}
  {\bibfnamefont {S.}~\bibnamefont {Paschen}}, \bibinfo {author} {\bibfnamefont
  {S.}~\bibnamefont {Kirchner}}, \ and\ \bibinfo {author} {\bibfnamefont
  {Q.}~\bibnamefont {Si}},\ }\href {\doibase 10.1073/pnas.1009202107}
  {\bibfield  {journal} {\bibinfo  {journal} {Proceedings of the National
  Academy of Sciences}\ }\textbf {\bibinfo {volume} {107}},\ \bibinfo {pages}
  {14547} (\bibinfo {year} {2010})}\BibitemShut {NoStop}%
\bibitem [{\citenamefont {Hartmann}\ \emph {et~al.}(2010)\citenamefont
  {Hartmann}, \citenamefont {Oeschler}, \citenamefont {Krellner}, \citenamefont
  {Geibel}, \citenamefont {Paschen},\ and\ \citenamefont
  {Steglich}}]{Hartmann2010}%
  \BibitemOpen
  \bibfield  {author} {\bibinfo {author} {\bibfnamefont {S.}~\bibnamefont
  {Hartmann}}, \bibinfo {author} {\bibfnamefont {N.}~\bibnamefont {Oeschler}},
  \bibinfo {author} {\bibfnamefont {C.}~\bibnamefont {Krellner}}, \bibinfo
  {author} {\bibfnamefont {C.}~\bibnamefont {Geibel}}, \bibinfo {author}
  {\bibfnamefont {S.}~\bibnamefont {Paschen}}, \ and\ \bibinfo {author}
  {\bibfnamefont {F.}~\bibnamefont {Steglich}},\ }\href {\doibase
  10.1103/PhysRevLett.104.096401} {\bibfield  {journal} {\bibinfo  {journal}
  {Phys. Rev. Lett.}\ }\textbf {\bibinfo {volume} {104}},\ \bibinfo {pages}
  {096401} (\bibinfo {year} {2010})}\BibitemShut {NoStop}%
\bibitem [{\citenamefont {Machida}\ \emph {et~al.}(2012)\citenamefont
  {Machida}, \citenamefont {Tomokuni}, \citenamefont {Ogura}, \citenamefont
  {Izawa}, \citenamefont {Kuga}, \citenamefont {Nakatsuji}, \citenamefont
  {Lapertot}, \citenamefont {Knebel}, \citenamefont {Brison},\ and\
  \citenamefont {Flouquet}}]{Machida2012a}%
  \BibitemOpen
  \bibfield  {author} {\bibinfo {author} {\bibfnamefont {Y.}~\bibnamefont
  {Machida}}, \bibinfo {author} {\bibfnamefont {K.}~\bibnamefont {Tomokuni}},
  \bibinfo {author} {\bibfnamefont {C.}~\bibnamefont {Ogura}}, \bibinfo
  {author} {\bibfnamefont {K.}~\bibnamefont {Izawa}}, \bibinfo {author}
  {\bibfnamefont {K.}~\bibnamefont {Kuga}}, \bibinfo {author} {\bibfnamefont
  {S.}~\bibnamefont {Nakatsuji}}, \bibinfo {author} {\bibfnamefont
  {G.}~\bibnamefont {Lapertot}}, \bibinfo {author} {\bibfnamefont
  {G.}~\bibnamefont {Knebel}}, \bibinfo {author} {\bibfnamefont {J.-P.}\
  \bibnamefont {Brison}}, \ and\ \bibinfo {author} {\bibfnamefont
  {J.}~\bibnamefont {Flouquet}},\ }\href {\doibase
  10.1103/PhysRevLett.109.156405} {\bibfield  {journal} {\bibinfo  {journal}
  {Phys. Rev. Lett.}\ }\textbf {\bibinfo {volume} {109}},\ \bibinfo {pages}
  {156405} (\bibinfo {year} {2012})}\BibitemShut {NoStop}%
\bibitem [{\citenamefont {Pfau}\ \emph {et~al.}(2012)\citenamefont {Pfau},
  \citenamefont {Hartmann}, \citenamefont {Stockert}, \citenamefont {Sun},
  \citenamefont {Lausberg}, \citenamefont {Brando}, \citenamefont {Friedemann},
  \citenamefont {Krellner}, \citenamefont {Geibel}, \citenamefont {Wirth},
  \citenamefont {Kirchner}, \citenamefont {Abrahams}, \citenamefont {Si},\ and\
  \citenamefont {Steglich}}]{Pfau2012}%
  \BibitemOpen
  \bibfield  {author} {\bibinfo {author} {\bibfnamefont {H.}~\bibnamefont
  {Pfau}}, \bibinfo {author} {\bibfnamefont {S.}~\bibnamefont {Hartmann}},
  \bibinfo {author} {\bibfnamefont {U.}~\bibnamefont {Stockert}}, \bibinfo
  {author} {\bibfnamefont {P.}~\bibnamefont {Sun}}, \bibinfo {author}
  {\bibfnamefont {S.}~\bibnamefont {Lausberg}}, \bibinfo {author}
  {\bibfnamefont {M.}~\bibnamefont {Brando}}, \bibinfo {author} {\bibfnamefont
  {S.}~\bibnamefont {Friedemann}}, \bibinfo {author} {\bibfnamefont
  {C.}~\bibnamefont {Krellner}}, \bibinfo {author} {\bibfnamefont
  {C.}~\bibnamefont {Geibel}}, \bibinfo {author} {\bibfnamefont
  {S.}~\bibnamefont {Wirth}}, \bibinfo {author} {\bibfnamefont
  {S.}~\bibnamefont {Kirchner}}, \bibinfo {author} {\bibfnamefont
  {E.}~\bibnamefont {Abrahams}}, \bibinfo {author} {\bibfnamefont
  {Q.}~\bibnamefont {Si}}, \ and\ \bibinfo {author} {\bibfnamefont
  {F.}~\bibnamefont {Steglich}},\ }\href
  {http://dx.doi.org/10.1038/nature11072} {\bibfield  {journal} {\bibinfo
  {journal} {Nature}\ }\textbf {\bibinfo {volume} {484}},\ \bibinfo {pages}
  {493} (\bibinfo {year} {2012})}\BibitemShut {NoStop}%
\bibitem [{\citenamefont {Machida}\ \emph {et~al.}(2013)\citenamefont
  {Machida}, \citenamefont {Tomokuni}, \citenamefont {Izawa}, \citenamefont
  {Lapertot}, \citenamefont {Knebel}, \citenamefont {Brison},\ and\
  \citenamefont {Flouquet}}]{Machida2013}%
  \BibitemOpen
  \bibfield  {author} {\bibinfo {author} {\bibfnamefont {Y.}~\bibnamefont
  {Machida}}, \bibinfo {author} {\bibfnamefont {K.}~\bibnamefont {Tomokuni}},
  \bibinfo {author} {\bibfnamefont {K.}~\bibnamefont {Izawa}}, \bibinfo
  {author} {\bibfnamefont {G.}~\bibnamefont {Lapertot}}, \bibinfo {author}
  {\bibfnamefont {G.}~\bibnamefont {Knebel}}, \bibinfo {author} {\bibfnamefont
  {J.-P.}\ \bibnamefont {Brison}}, \ and\ \bibinfo {author} {\bibfnamefont
  {J.}~\bibnamefont {Flouquet}},\ }\href {\doibase
  10.1103/PhysRevLett.110.236402} {\bibfield  {journal} {\bibinfo  {journal}
  {Phys. Rev. Lett.}\ }\textbf {\bibinfo {volume} {110}},\ \bibinfo {pages}
  {236402} (\bibinfo {year} {2013})}\BibitemShut {NoStop}%
\bibitem [{\citenamefont {Reid}\ \emph {et~al.}(2014)\citenamefont {Reid},
  \citenamefont {Tanatar}, \citenamefont {Daou}, \citenamefont {Hu},
  \citenamefont {Petrovic},\ and\ \citenamefont {Taillefer}}]{Reid2014}%
  \BibitemOpen
  \bibfield  {author} {\bibinfo {author} {\bibfnamefont {J.-P.}\ \bibnamefont
  {Reid}}, \bibinfo {author} {\bibfnamefont {M.~A.}\ \bibnamefont {Tanatar}},
  \bibinfo {author} {\bibfnamefont {R.}~\bibnamefont {Daou}}, \bibinfo {author}
  {\bibfnamefont {R.}~\bibnamefont {Hu}}, \bibinfo {author} {\bibfnamefont
  {C.}~\bibnamefont {Petrovic}}, \ and\ \bibinfo {author} {\bibfnamefont
  {L.}~\bibnamefont {Taillefer}},\ }\href {\doibase 10.1103/PhysRevB.89.045130}
  {\bibfield  {journal} {\bibinfo  {journal} {Phys. Rev. B}\ }\textbf {\bibinfo
  {volume} {89}},\ \bibinfo {pages} {045130} (\bibinfo {year}
  {2014})}\BibitemShut {NoStop}%
\bibitem [{\citenamefont {Paglione}\ \emph {et~al.}(2003)\citenamefont
  {Paglione}, \citenamefont {Tanatar}, \citenamefont {Hawthorn}, \citenamefont
  {Boaknin}, \citenamefont {Hill}, \citenamefont {Ronning}, \citenamefont
  {Sutherland},\ and\ \citenamefont {Taillefer}}]{Paglione2003}%
  \BibitemOpen
  \bibfield  {author} {\bibinfo {author} {\bibfnamefont {J.}~\bibnamefont
  {Paglione}}, \bibinfo {author} {\bibfnamefont {M.~A.}\ \bibnamefont
  {Tanatar}}, \bibinfo {author} {\bibfnamefont {D.}~\bibnamefont {Hawthorn}},
  \bibinfo {author} {\bibfnamefont {E.}~\bibnamefont {Boaknin}}, \bibinfo
  {author} {\bibfnamefont {R.}~\bibnamefont {Hill}}, \bibinfo {author}
  {\bibfnamefont {F.}~\bibnamefont {Ronning}}, \bibinfo {author} {\bibfnamefont
  {M.}~\bibnamefont {Sutherland}}, \ and\ \bibinfo {author} {\bibfnamefont
  {L.}~\bibnamefont {Taillefer}},\ }\href {\doibase
  10.1103/PhysRevLett.91.246405} {\bibfield  {journal} {\bibinfo  {journal}
  {Phys. Rev. Lett.}\ }\textbf {\bibinfo {volume} {91}},\ \bibinfo {pages}
  {246405} (\bibinfo {year} {2003})}\BibitemShut {NoStop}%
\bibitem [{\citenamefont {Pourret}\ \emph {et~al.}(2013)\citenamefont
  {Pourret}, \citenamefont {Knebel}, \citenamefont {Matsuda}, \citenamefont
  {Lapertot},\ and\ \citenamefont {Flouquet}}]{Pourret2013}%
  \BibitemOpen
  \bibfield  {author} {\bibinfo {author} {\bibfnamefont {A.}~\bibnamefont
  {Pourret}}, \bibinfo {author} {\bibfnamefont {G.}~\bibnamefont {Knebel}},
  \bibinfo {author} {\bibfnamefont {T.~D.}\ \bibnamefont {Matsuda}}, \bibinfo
  {author} {\bibfnamefont {G.}~\bibnamefont {Lapertot}}, \ and\ \bibinfo
  {author} {\bibfnamefont {J.}~\bibnamefont {Flouquet}},\ }\href {\doibase
  10.7566/JPSJ.82.053704} {\bibfield  {journal} {\bibinfo  {journal} {Journal
  of the Physical Society of Japan}\ }\textbf {\bibinfo {volume} {82}},\
  \bibinfo {pages} {053704} (\bibinfo {year} {2013})}\BibitemShut {NoStop}%
\bibitem [{\citenamefont {Schuberth}\ \emph {et~al.}(2009)\citenamefont
  {Schuberth}, \citenamefont {Tippmann}, \citenamefont {Kath}, \citenamefont
  {Krellner}, \citenamefont {Geibel}, \citenamefont {Westerkamp}, \citenamefont
  {Klingner},\ and\ \citenamefont {Steglich}}]{Schuberth2009}%
  \BibitemOpen
  \bibfield  {author} {\bibinfo {author} {\bibfnamefont {E.}~\bibnamefont
  {Schuberth}}, \bibinfo {author} {\bibfnamefont {M.}~\bibnamefont {Tippmann}},
  \bibinfo {author} {\bibfnamefont {M.}~\bibnamefont {Kath}}, \bibinfo {author}
  {\bibfnamefont {C.}~\bibnamefont {Krellner}}, \bibinfo {author}
  {\bibfnamefont {C.}~\bibnamefont {Geibel}}, \bibinfo {author} {\bibfnamefont
  {T.}~\bibnamefont {Westerkamp}}, \bibinfo {author} {\bibfnamefont
  {C.}~\bibnamefont {Klingner}}, \ and\ \bibinfo {author} {\bibfnamefont
  {F.}~\bibnamefont {Steglich}},\ }\href
  {http://stacks.iop.org/1742-6596/150/i=4/a=042178} {\bibfield  {journal}
  {\bibinfo  {journal} {Journal of Physics: Conference Series}\ }\textbf
  {\bibinfo {volume} {150}},\ \bibinfo {pages} {042178} (\bibinfo {year}
  {2009})}\BibitemShut {NoStop}%
\bibitem [{\citenamefont {Schuberth}\ \emph {et~al.}(2013)\citenamefont
  {Schuberth}, \citenamefont {Tippmann}, \citenamefont {Krellner},\ and\
  \citenamefont {Steglich}}]{Schuberth2013}%
  \BibitemOpen
  \bibfield  {author} {\bibinfo {author} {\bibfnamefont {E.}~\bibnamefont
  {Schuberth}}, \bibinfo {author} {\bibfnamefont {M.}~\bibnamefont {Tippmann}},
  \bibinfo {author} {\bibfnamefont {C.}~\bibnamefont {Krellner}}, \ and\
  \bibinfo {author} {\bibfnamefont {F.}~\bibnamefont {Steglich}},\ }\href
  {\doibase 10.1002/pssb.201200786} {\bibfield  {journal} {\bibinfo  {journal}
  {Physica Status Solidi (B)}\ }\textbf {\bibinfo {volume} {250}},\ \bibinfo
  {pages} {482} (\bibinfo {year} {2013})}\BibitemShut {NoStop}%
\bibitem [{\citenamefont {Taupin}\ \emph {et~al.}(2014)\citenamefont {Taupin},
  \citenamefont {Howald}, \citenamefont {Aoki}, \citenamefont {Flouquet},\ and\
  \citenamefont {Brison}}]{Taupin2014}%
  \BibitemOpen
  \bibfield  {author} {\bibinfo {author} {\bibfnamefont {M.}~\bibnamefont
  {Taupin}}, \bibinfo {author} {\bibfnamefont {L.}~\bibnamefont {Howald}},
  \bibinfo {author} {\bibfnamefont {D.}~\bibnamefont {Aoki}}, \bibinfo {author}
  {\bibfnamefont {J.}~\bibnamefont {Flouquet}}, \ and\ \bibinfo {author}
  {\bibfnamefont {J.~P.}\ \bibnamefont {Brison}},\ }\href {\doibase
  10.1103/PhysRevB.89.041108} {\bibfield  {journal} {\bibinfo  {journal} {Phys.
  Rev. B}\ }\textbf {\bibinfo {volume} {89}},\ \bibinfo {pages} {041108}
  (\bibinfo {year} {2014})}\BibitemShut {NoStop}%
\bibitem [{\citenamefont {Wagner}\ \emph {et~al.}(1971)\citenamefont {Wagner},
  \citenamefont {Garland},\ and\ \citenamefont {Bowers}}]{Wagner1971}%
  \BibitemOpen
  \bibfield  {author} {\bibinfo {author} {\bibfnamefont {D.~K.}\ \bibnamefont
  {Wagner}}, \bibinfo {author} {\bibfnamefont {J.~C.}\ \bibnamefont {Garland}},
  \ and\ \bibinfo {author} {\bibfnamefont {R.}~\bibnamefont {Bowers}},\ }\href
  {\doibase 10.1103/PhysRevB.3.3141} {\bibfield  {journal} {\bibinfo  {journal}
  {Phys. Rev. B}\ }\textbf {\bibinfo {volume} {3}},\ \bibinfo {pages} {3141}
  (\bibinfo {year} {1971})}\BibitemShut {NoStop}%
\bibitem [{\citenamefont {Lussier}\ \emph {et~al.}(1994)\citenamefont
  {Lussier}, \citenamefont {Ellman},\ and\ \citenamefont
  {Taillefer}}]{Lussier1994}%
  \BibitemOpen
  \bibfield  {author} {\bibinfo {author} {\bibfnamefont {B.}~\bibnamefont
  {Lussier}}, \bibinfo {author} {\bibfnamefont {B.}~\bibnamefont {Ellman}}, \
  and\ \bibinfo {author} {\bibfnamefont {L.}~\bibnamefont {Taillefer}},\ }\href
  {\doibase 10.1103/PhysRevLett.73.3294} {\bibfield  {journal} {\bibinfo
  {journal} {Phys. Rev. Lett.}\ }\textbf {\bibinfo {volume} {73}},\ \bibinfo
  {pages} {3294} (\bibinfo {year} {1994})}\BibitemShut {NoStop}%
\bibitem [{\citenamefont {Paglione}\ \emph {et~al.}(2005)\citenamefont
  {Paglione}, \citenamefont {Tanatar}, \citenamefont {Hawthorn}, \citenamefont
  {Hill}, \citenamefont {Ronning}, \citenamefont {Sutherland}, \citenamefont
  {Taillefer}, \citenamefont {Petrovic},\ and\ \citenamefont
  {Canfield}}]{Paglione2005}%
  \BibitemOpen
  \bibfield  {author} {\bibinfo {author} {\bibfnamefont {J.}~\bibnamefont
  {Paglione}}, \bibinfo {author} {\bibfnamefont {M.~A.}\ \bibnamefont
  {Tanatar}}, \bibinfo {author} {\bibfnamefont {D.~G.}\ \bibnamefont
  {Hawthorn}}, \bibinfo {author} {\bibfnamefont {R.~W.}\ \bibnamefont {Hill}},
  \bibinfo {author} {\bibfnamefont {F.}~\bibnamefont {Ronning}}, \bibinfo
  {author} {\bibfnamefont {M.}~\bibnamefont {Sutherland}}, \bibinfo {author}
  {\bibfnamefont {L.}~\bibnamefont {Taillefer}}, \bibinfo {author}
  {\bibfnamefont {C.}~\bibnamefont {Petrovic}}, \ and\ \bibinfo {author}
  {\bibfnamefont {P.~C.}\ \bibnamefont {Canfield}},\ }\href {\doibase
  10.1103/PhysRevLett.94.216602} {\bibfield  {journal} {\bibinfo  {journal}
  {Phys. Rev. Lett.}\ }\textbf {\bibinfo {volume} {94}},\ \bibinfo {pages}
  {216602} (\bibinfo {year} {2005})}\BibitemShut {NoStop}%
\bibitem [{\citenamefont {Ishida}\ \emph {et~al.}(2002)\citenamefont {Ishida},
  \citenamefont {Okamoto}, \citenamefont {Kawasaki}, \citenamefont {Kitaoka},
  \citenamefont {Trovarelli}, \citenamefont {Geibel},\ and\ \citenamefont
  {Steglich}}]{Ishida2002}%
  \BibitemOpen
  \bibfield  {author} {\bibinfo {author} {\bibfnamefont {K.}~\bibnamefont
  {Ishida}}, \bibinfo {author} {\bibfnamefont {K.}~\bibnamefont {Okamoto}},
  \bibinfo {author} {\bibfnamefont {Y.}~\bibnamefont {Kawasaki}}, \bibinfo
  {author} {\bibfnamefont {Y.}~\bibnamefont {Kitaoka}}, \bibinfo {author}
  {\bibfnamefont {O.}~\bibnamefont {Trovarelli}}, \bibinfo {author}
  {\bibfnamefont {C.}~\bibnamefont {Geibel}}, \ and\ \bibinfo {author}
  {\bibfnamefont {F.}~\bibnamefont {Steglich}},\ }\href {\doibase
  10.1103/PhysRevLett.89.107202} {\bibfield  {journal} {\bibinfo  {journal}
  {Phys. Rev. Lett.}\ }\textbf {\bibinfo {volume} {89}},\ \bibinfo {pages}
  {107202} (\bibinfo {year} {2002})}\BibitemShut {NoStop}%
\bibitem [{\citenamefont {Kambe}\ \emph {et~al.}(2014)\citenamefont {Kambe},
  \citenamefont {Sakai}, \citenamefont {Tokunaga}, \citenamefont {Lapertot},
  \citenamefont {Matsuda}, \citenamefont {Knebel}, \citenamefont {Flouquet},\
  and\ \citenamefont {Walstedt}}]{Kambe2014a}%
  \BibitemOpen
  \bibfield  {author} {\bibinfo {author} {\bibfnamefont {S.}~\bibnamefont
  {Kambe}}, \bibinfo {author} {\bibfnamefont {H.}~\bibnamefont {Sakai}},
  \bibinfo {author} {\bibfnamefont {Y.}~\bibnamefont {Tokunaga}}, \bibinfo
  {author} {\bibfnamefont {G.}~\bibnamefont {Lapertot}}, \bibinfo {author}
  {\bibfnamefont {T.~D.}\ \bibnamefont {Matsuda}}, \bibinfo {author}
  {\bibfnamefont {G.}~\bibnamefont {Knebel}}, \bibinfo {author} {\bibfnamefont
  {J.}~\bibnamefont {Flouquet}}, \ and\ \bibinfo {author} {\bibfnamefont
  {R.~E.}\ \bibnamefont {Walstedt}},\ }\href
  {http://dx.doi.org/10.1038/nphys3101} {\bibfield  {journal} {\bibinfo
  {journal} {Nat Phys}\ }\textbf {\bibinfo {volume} {10}},\ \bibinfo {pages}
  {840} (\bibinfo {year} {2014})}\BibitemShut {NoStop}%
\bibitem [{\citenamefont {Stock}\ \emph {et~al.}(2012)\citenamefont {Stock},
  \citenamefont {Broholm}, \citenamefont {Demmel}, \citenamefont {Van~Duijn},
  \citenamefont {Taylor}, \citenamefont {Kang}, \citenamefont {Hu},\ and\
  \citenamefont {Petrovic}}]{Stock2012}%
  \BibitemOpen
  \bibfield  {author} {\bibinfo {author} {\bibfnamefont {C.}~\bibnamefont
  {Stock}}, \bibinfo {author} {\bibfnamefont {C.}~\bibnamefont {Broholm}},
  \bibinfo {author} {\bibfnamefont {F.}~\bibnamefont {Demmel}}, \bibinfo
  {author} {\bibfnamefont {J.}~\bibnamefont {Van~Duijn}}, \bibinfo {author}
  {\bibfnamefont {J.~W.}\ \bibnamefont {Taylor}}, \bibinfo {author}
  {\bibfnamefont {H.~J.}\ \bibnamefont {Kang}}, \bibinfo {author}
  {\bibfnamefont {R.}~\bibnamefont {Hu}}, \ and\ \bibinfo {author}
  {\bibfnamefont {C.}~\bibnamefont {Petrovic}},\ }\href {\doibase
  10.1103/PhysRevLett.109.127201} {\bibfield  {journal} {\bibinfo  {journal}
  {Phys. Rev. Lett.}\ }\textbf {\bibinfo {volume} {109}},\ \bibinfo {pages}
  {127201} (\bibinfo {year} {2012})}\BibitemShut {NoStop}%
\bibitem [{\citenamefont {Sichelschmidt}\ \emph {et~al.}(2003)\citenamefont
  {Sichelschmidt}, \citenamefont {Ivanshin}, \citenamefont {Ferstl},
  \citenamefont {Geibel},\ and\ \citenamefont {Steglich}}]{Sichelschmidt2003}%
  \BibitemOpen
  \bibfield  {author} {\bibinfo {author} {\bibfnamefont {J.}~\bibnamefont
  {Sichelschmidt}}, \bibinfo {author} {\bibfnamefont {V.~A.}\ \bibnamefont
  {Ivanshin}}, \bibinfo {author} {\bibfnamefont {J.}~\bibnamefont {Ferstl}},
  \bibinfo {author} {\bibfnamefont {C.}~\bibnamefont {Geibel}}, \ and\ \bibinfo
  {author} {\bibfnamefont {F.}~\bibnamefont {Steglich}},\ }\href {\doibase
  10.1103/PhysRevLett.91.156401} {\bibfield  {journal} {\bibinfo  {journal}
  {Phys. Rev. Lett.}\ }\textbf {\bibinfo {volume} {91}},\ \bibinfo {pages}
  {156401} (\bibinfo {year} {2003})}\BibitemShut {NoStop}%
\bibitem [{\citenamefont {Abrahams}\ and\ \citenamefont
  {W\"olfle}(2008)}]{Abrahams2008}%
  \BibitemOpen
  \bibfield  {author} {\bibinfo {author} {\bibfnamefont {E.}~\bibnamefont
  {Abrahams}}\ and\ \bibinfo {author} {\bibfnamefont {P.}~\bibnamefont
  {W\"olfle}},\ }\href {\doibase 10.1103/PhysRevB.78.104423} {\bibfield
  {journal} {\bibinfo  {journal} {Phys. Rev. B}\ }\textbf {\bibinfo {volume}
  {78}},\ \bibinfo {pages} {104423} (\bibinfo {year} {2008})}\BibitemShut
  {NoStop}%
\bibitem [{\citenamefont {Knebel}\ \emph {et~al.}(2006)\citenamefont {Knebel},
  \citenamefont {Aoki}, \citenamefont {Braithwaite}, \citenamefont {Salce},\
  and\ \citenamefont {Flouquet}}]{Knebel2006a}%
  \BibitemOpen
  \bibfield  {author} {\bibinfo {author} {\bibfnamefont {G.}~\bibnamefont
  {Knebel}}, \bibinfo {author} {\bibfnamefont {D.}~\bibnamefont {Aoki}},
  \bibinfo {author} {\bibfnamefont {D.}~\bibnamefont {Braithwaite}}, \bibinfo
  {author} {\bibfnamefont {B.}~\bibnamefont {Salce}}, \ and\ \bibinfo {author}
  {\bibfnamefont {J.}~\bibnamefont {Flouquet}},\ }\href {\doibase
  10.1103/PhysRevB.74.020501} {\bibfield  {journal} {\bibinfo  {journal} {Phys.
  Rev. B}\ }\textbf {\bibinfo {volume} {74}},\ \bibinfo {pages} {020501}
  (\bibinfo {year} {2006})}\BibitemShut {NoStop}%
\bibitem [{\citenamefont {Lausberg}\ \emph {et~al.}(2013)\citenamefont
  {Lausberg}, \citenamefont {Hannaske}, \citenamefont {Steppke}, \citenamefont
  {Steinke}, \citenamefont {Gruner}, \citenamefont {Pedrero}, \citenamefont
  {Krellner}, \citenamefont {Klingner}, \citenamefont {Brando}, \citenamefont
  {Geibel},\ and\ \citenamefont {Steglich}}]{Lausberg2013}%
  \BibitemOpen
  \bibfield  {author} {\bibinfo {author} {\bibfnamefont {S.}~\bibnamefont
  {Lausberg}}, \bibinfo {author} {\bibfnamefont {A.}~\bibnamefont {Hannaske}},
  \bibinfo {author} {\bibfnamefont {A.}~\bibnamefont {Steppke}}, \bibinfo
  {author} {\bibfnamefont {L.}~\bibnamefont {Steinke}}, \bibinfo {author}
  {\bibfnamefont {T.}~\bibnamefont {Gruner}}, \bibinfo {author} {\bibfnamefont
  {L.}~\bibnamefont {Pedrero}}, \bibinfo {author} {\bibfnamefont
  {C.}~\bibnamefont {Krellner}}, \bibinfo {author} {\bibfnamefont
  {C.}~\bibnamefont {Klingner}}, \bibinfo {author} {\bibfnamefont
  {M.}~\bibnamefont {Brando}}, \bibinfo {author} {\bibfnamefont
  {C.}~\bibnamefont {Geibel}}, \ and\ \bibinfo {author} {\bibfnamefont
  {F.}~\bibnamefont {Steglich}},\ }\href {\doibase
  10.1103/PhysRevLett.110.256402} {\bibfield  {journal} {\bibinfo  {journal}
  {Phys. Rev. Lett.}\ }\textbf {\bibinfo {volume} {110}},\ \bibinfo {pages}
  {256402} (\bibinfo {year} {2013})}\BibitemShut {NoStop}%
\bibitem [{\citenamefont {Ueda}\ and\ \citenamefont {Moriya}(1975)}]{Ueda1975}%
  \BibitemOpen
  \bibfield  {author} {\bibinfo {author} {\bibfnamefont {K.}~\bibnamefont
  {Ueda}}\ and\ \bibinfo {author} {\bibfnamefont {T.}~\bibnamefont {Moriya}},\
  }\href {\doibase 10.1143/JPSJ.39.605} {\bibfield  {journal} {\bibinfo
  {journal} {Journal of the Physical Society of Japan}\ }\textbf {\bibinfo
  {volume} {39}},\ \bibinfo {pages} {605} (\bibinfo {year} {1975})}\BibitemShut
  {NoStop}%
\bibitem [{\citenamefont {G\"uttler}\ \emph {et~al.}(2014)\citenamefont
  {G\"uttler}, \citenamefont {Kummer}, \citenamefont {Patil}, \citenamefont
  {H\"oppner}, \citenamefont {Hannaske}, \citenamefont {Danzenb\"acher},
  \citenamefont {Shi}, \citenamefont {Radovic}, \citenamefont {Rienks},
  \citenamefont {Laubschat}, \citenamefont {Geibel},\ and\ \citenamefont
  {Vyalikh}}]{Guttler2014}%
  \BibitemOpen
  \bibfield  {author} {\bibinfo {author} {\bibfnamefont {M.}~\bibnamefont
  {G\"uttler}}, \bibinfo {author} {\bibfnamefont {K.}~\bibnamefont {Kummer}},
  \bibinfo {author} {\bibfnamefont {S.}~\bibnamefont {Patil}}, \bibinfo
  {author} {\bibfnamefont {M.}~\bibnamefont {H\"oppner}}, \bibinfo {author}
  {\bibfnamefont {A.}~\bibnamefont {Hannaske}}, \bibinfo {author}
  {\bibfnamefont {S.}~\bibnamefont {Danzenb\"acher}}, \bibinfo {author}
  {\bibfnamefont {M.}~\bibnamefont {Shi}}, \bibinfo {author} {\bibfnamefont
  {M.}~\bibnamefont {Radovic}}, \bibinfo {author} {\bibfnamefont
  {E.}~\bibnamefont {Rienks}}, \bibinfo {author} {\bibfnamefont
  {C.}~\bibnamefont {Laubschat}}, \bibinfo {author} {\bibfnamefont
  {C.}~\bibnamefont {Geibel}}, \ and\ \bibinfo {author} {\bibfnamefont {D.~V.}\
  \bibnamefont {Vyalikh}},\ }\href {\doibase 10.1103/PhysRevB.90.195138}
  {\bibfield  {journal} {\bibinfo  {journal} {Phys. Rev. B}\ }\textbf {\bibinfo
  {volume} {90}},\ \bibinfo {pages} {195138} (\bibinfo {year}
  {2014})}\BibitemShut {NoStop}%
\bibitem [{\citenamefont {Danzenb\"acher}\ \emph {et~al.}(2011)\citenamefont
  {Danzenb\"acher}, \citenamefont {Vyalikh}, \citenamefont {Kummer},
  \citenamefont {Krellner}, \citenamefont {Holder}, \citenamefont {H\"oppner},
  \citenamefont {Kucherenko}, \citenamefont {Geibel}, \citenamefont {Shi},
  \citenamefont {Patthey}, \citenamefont {Molodtsov},\ and\ \citenamefont
  {Laubschat}}]{Danzenbacher2011}%
  \BibitemOpen
  \bibfield  {author} {\bibinfo {author} {\bibfnamefont {S.}~\bibnamefont
  {Danzenb\"acher}}, \bibinfo {author} {\bibfnamefont {D.~V.}\ \bibnamefont
  {Vyalikh}}, \bibinfo {author} {\bibfnamefont {K.}~\bibnamefont {Kummer}},
  \bibinfo {author} {\bibfnamefont {C.}~\bibnamefont {Krellner}}, \bibinfo
  {author} {\bibfnamefont {M.}~\bibnamefont {Holder}}, \bibinfo {author}
  {\bibfnamefont {M.}~\bibnamefont {H\"oppner}}, \bibinfo {author}
  {\bibfnamefont {Y.}~\bibnamefont {Kucherenko}}, \bibinfo {author}
  {\bibfnamefont {C.}~\bibnamefont {Geibel}}, \bibinfo {author} {\bibfnamefont
  {M.}~\bibnamefont {Shi}}, \bibinfo {author} {\bibfnamefont {L.}~\bibnamefont
  {Patthey}}, \bibinfo {author} {\bibfnamefont {S.~L.}\ \bibnamefont
  {Molodtsov}}, \ and\ \bibinfo {author} {\bibfnamefont {C.}~\bibnamefont
  {Laubschat}},\ }\href {\doibase 10.1103/PhysRevLett.107.267601} {\bibfield
  {journal} {\bibinfo  {journal} {Phys. Rev. Lett.}\ }\textbf {\bibinfo
  {volume} {107}},\ \bibinfo {pages} {267601} (\bibinfo {year}
  {2011})}\BibitemShut {NoStop}%
\bibitem [{\citenamefont {Abrahams}\ \emph {et~al.}(2014)\citenamefont
  {Abrahams}, \citenamefont {Schmalian},\ and\ \citenamefont
  {W\"olfle}}]{Abrahams2014}%
  \BibitemOpen
  \bibfield  {author} {\bibinfo {author} {\bibfnamefont {E.}~\bibnamefont
  {Abrahams}}, \bibinfo {author} {\bibfnamefont {J.}~\bibnamefont {Schmalian}},
  \ and\ \bibinfo {author} {\bibfnamefont {P.}~\bibnamefont {W\"olfle}},\
  }\href {\doibase 10.1103/PhysRevB.90.045105} {\bibfield  {journal} {\bibinfo
  {journal} {Phys. Rev. B}\ }\textbf {\bibinfo {volume} {90}},\ \bibinfo
  {pages} {045105} (\bibinfo {year} {2014})}\BibitemShut {NoStop}%
\end{thebibliography}

\begin{thebibliography}{6}%
\makeatletter
\providecommand \@ifxundefined [1]{%
 \@ifx{#1\undefined}
}%
\providecommand \@ifnum [1]{%
 \ifnum #1\expandafter \@firstoftwo
 \else \expandafter \@secondoftwo
 \fi
}%
\providecommand \@ifx [1]{%
 \ifx #1\expandafter \@firstoftwo
 \else \expandafter \@secondoftwo
 \fi
}%
\providecommand \natexlab [1]{#1}%
\providecommand \enquote  [1]{``#1''}%
\providecommand \bibnamefont  [1]{#1}%
\providecommand \bibfnamefont [1]{#1}%
\providecommand \citenamefont [1]{#1}%
\providecommand \href@noop [0]{\@secondoftwo}%
\providecommand \href [0]{\begingroup \@sanitize@url \@href}%
\providecommand \@href[1]{\@@startlink{#1}\@@href}%
\providecommand \@@href[1]{\endgroup#1\@@endlink}%
\providecommand \@sanitize@url [0]{\catcode `\\12\catcode `\$12\catcode
  `\&12\catcode `\#12\catcode `\^12\catcode `\_12\catcode `\%12\relax}%
\providecommand \@@startlink[1]{}%
\providecommand \@@endlink[0]{}%
\providecommand \url  [0]{\begingroup\@sanitize@url \@url }%
\providecommand \@url [1]{\endgroup\@href {#1}{\urlprefix }}%
\providecommand \urlprefix  [0]{URL }%
\providecommand \Eprint [0]{\href }%
\providecommand \doibase [0]{http://dx.doi.org/}%
\providecommand \selectlanguage [0]{\@gobble}%
\providecommand \bibinfo  [0]{\@secondoftwo}%
\providecommand \bibfield  [0]{\@secondoftwo}%
\providecommand \translation [1]{[#1]}%
\providecommand \BibitemOpen [0]{}%
\providecommand \bibitemStop [0]{}%
\providecommand \bibitemNoStop [0]{.\EOS\space}%
\providecommand \EOS [0]{\spacefactor3000\relax}%
\providecommand \BibitemShut  [1]{\csname bibitem#1\endcsname}%
\let\auto@bib@innerbib\@empty
\bibitem [{\citenamefont {Kambe}\ \emph {et~al.}(2014)\citenamefont {Kambe},
  \citenamefont {Sakai}, \citenamefont {Tokunaga}, \citenamefont {Lapertot},
  \citenamefont {Matsuda}, \citenamefont {Knebel}, \citenamefont {Flouquet},\
  and\ \citenamefont {Walstedt}}]{Kambe2014a}%
  \BibitemOpen
  \bibfield  {author} {\bibinfo {author} {\bibfnamefont {S.}~\bibnamefont
  {Kambe}}, \bibinfo {author} {\bibfnamefont {H.}~\bibnamefont {Sakai}},
  \bibinfo {author} {\bibfnamefont {Y.}~\bibnamefont {Tokunaga}}, \bibinfo
  {author} {\bibfnamefont {G.}~\bibnamefont {Lapertot}}, \bibinfo {author}
  {\bibfnamefont {T.~D.}\ \bibnamefont {Matsuda}}, \bibinfo {author}
  {\bibfnamefont {G.}~\bibnamefont {Knebel}}, \bibinfo {author} {\bibfnamefont
  {J.}~\bibnamefont {Flouquet}}, \ and\ \bibinfo {author} {\bibfnamefont
  {R.~E.}\ \bibnamefont {Walstedt}},\ }\href
  {http://dx.doi.org/10.1038/nphys3101} {\bibfield  {journal} {\bibinfo
  {journal} {Nat. Phys.}\ }\textbf {\bibinfo {volume} {10}},\ \bibinfo {pages}
  {840} (\bibinfo {year} {2014})}\BibitemShut {NoStop}%
\bibitem [{\citenamefont {Stock}\ \emph {et~al.}(2012)\citenamefont {Stock},
  \citenamefont {Broholm}, \citenamefont {Demmel}, \citenamefont {Van~Duijn},
  \citenamefont {Taylor}, \citenamefont {Kang}, \citenamefont {Hu},\ and\
  \citenamefont {Petrovic}}]{Stock2012}%
  \BibitemOpen
  \bibfield  {author} {\bibinfo {author} {\bibfnamefont {C.}~\bibnamefont
  {Stock}}, \bibinfo {author} {\bibfnamefont {C.}~\bibnamefont {Broholm}},
  \bibinfo {author} {\bibfnamefont {F.}~\bibnamefont {Demmel}}, \bibinfo
  {author} {\bibfnamefont {J.}~\bibnamefont {Van~Duijn}}, \bibinfo {author}
  {\bibfnamefont {J.~W.}\ \bibnamefont {Taylor}}, \bibinfo {author}
  {\bibfnamefont {H.~J.}\ \bibnamefont {Kang}}, \bibinfo {author}
  {\bibfnamefont {R.}~\bibnamefont {Hu}}, \ and\ \bibinfo {author}
  {\bibfnamefont {C.}~\bibnamefont {Petrovic}},\ }\href {\doibase
  10.1103/PhysRevLett.109.127201} {\bibfield  {journal} {\bibinfo  {journal}
  {Phys. Rev. Lett.}\ }\textbf {\bibinfo {volume} {109}},\ \bibinfo {pages}
  {127201} (\bibinfo {year} {2012})}\BibitemShut {NoStop}%
\bibitem [{\citenamefont {Ueda}\ and\ \citenamefont {Moriya}(1975)}]{Ueda1975}%
  \BibitemOpen
  \bibfield  {author} {\bibinfo {author} {\bibfnamefont {K.}~\bibnamefont
  {Ueda}}\ and\ \bibinfo {author} {\bibfnamefont {T.}~\bibnamefont {Moriya}},\
  }\href {\doibase 10.1143/JPSJ.39.605} {\bibfield  {journal} {\bibinfo
  {journal} {J. Phys. Soc. of Jpn}\ }\textbf {\bibinfo {volume} {39}},\
  \bibinfo {pages} {605} (\bibinfo {year} {1975})}\BibitemShut {NoStop}%
\bibitem [{\citenamefont {Smith}\ \emph {et~al.}(2008)\citenamefont {Smith},
  \citenamefont {Sutherland}, \citenamefont {Lonzarich}, \citenamefont
  {Saxena}, \citenamefont {Kimura}, \citenamefont {Takashima}, \citenamefont
  {Nohara},\ and\ \citenamefont {Takagi}}]{Smith2008}%
  \BibitemOpen
  \bibfield  {author} {\bibinfo {author} {\bibfnamefont {R.~P.}\ \bibnamefont
  {Smith}}, \bibinfo {author} {\bibfnamefont {M.}~\bibnamefont {Sutherland}},
  \bibinfo {author} {\bibfnamefont {G.~G.}\ \bibnamefont {Lonzarich}}, \bibinfo
  {author} {\bibfnamefont {S.~S.}\ \bibnamefont {Saxena}}, \bibinfo {author}
  {\bibfnamefont {N.}~\bibnamefont {Kimura}}, \bibinfo {author} {\bibfnamefont
  {S.}~\bibnamefont {Takashima}}, \bibinfo {author} {\bibfnamefont
  {M.}~\bibnamefont {Nohara}}, \ and\ \bibinfo {author} {\bibfnamefont
  {H.}~\bibnamefont {Takagi}},\ }\href {http://dx.doi.org/10.1038/nature07401}
  {\bibfield  {journal} {\bibinfo  {journal} {Nature}\ }\textbf {\bibinfo
  {volume} {455}},\ \bibinfo {pages} {1220} (\bibinfo {year}
  {2008})}\BibitemShut {NoStop}%
\bibitem [{\citenamefont {Pfau}\ \emph {et~al.}(2012)\citenamefont {Pfau},
  \citenamefont {Hartmann}, \citenamefont {Stockert}, \citenamefont {Sun},
  \citenamefont {Lausberg}, \citenamefont {Brando}, \citenamefont {Friedemann},
  \citenamefont {Krellner}, \citenamefont {Geibel}, \citenamefont {Wirth},
  \citenamefont {Kirchner}, \citenamefont {Abrahams}, \citenamefont {Si},\ and\
  \citenamefont {Steglich}}]{Pfau2012}%
  \BibitemOpen
  \bibfield  {author} {\bibinfo {author} {\bibfnamefont {H.}~\bibnamefont
  {Pfau}}, \bibinfo {author} {\bibfnamefont {S.}~\bibnamefont {Hartmann}},
  \bibinfo {author} {\bibfnamefont {U.}~\bibnamefont {Stockert}}, \bibinfo
  {author} {\bibfnamefont {P.}~\bibnamefont {Sun}}, \bibinfo {author}
  {\bibfnamefont {S.}~\bibnamefont {Lausberg}}, \bibinfo {author}
  {\bibfnamefont {M.}~\bibnamefont {Brando}}, \bibinfo {author} {\bibfnamefont
  {S.}~\bibnamefont {Friedemann}}, \bibinfo {author} {\bibfnamefont
  {C.}~\bibnamefont {Krellner}}, \bibinfo {author} {\bibfnamefont
  {C.}~\bibnamefont {Geibel}}, \bibinfo {author} {\bibfnamefont
  {S.}~\bibnamefont {Wirth}}, \bibinfo {author} {\bibfnamefont
  {S.}~\bibnamefont {Kirchner}}, \bibinfo {author} {\bibfnamefont
  {E.}~\bibnamefont {Abrahams}}, \bibinfo {author} {\bibfnamefont
  {Q.}~\bibnamefont {Si}}, \ and\ \bibinfo {author} {\bibfnamefont
  {F.}~\bibnamefont {Steglich}},\ }\href
  {http://dx.doi.org/10.1038/nature11072} {\bibfield  {journal} {\bibinfo
  {journal} {Nature}\ }\textbf {\bibinfo {volume} {484}},\ \bibinfo {pages}
  {493} (\bibinfo {year} {2012})}\BibitemShut {NoStop}%
\bibitem [{\citenamefont {Flouquet}\ \emph {et~al.}(2004)\citenamefont
  {Flouquet}, \citenamefont {Haga}, \citenamefont {Haen}, \citenamefont
  {Braithwaite}, \citenamefont {Knebel}, \citenamefont {Raymond},\ and\
  \citenamefont {Kambe}}]{Flouquet2004}%
  \BibitemOpen
  \bibfield  {author} {\bibinfo {author} {\bibfnamefont {J.}~\bibnamefont
  {Flouquet}}, \bibinfo {author} {\bibfnamefont {Y.}~\bibnamefont {Haga}},
  \bibinfo {author} {\bibfnamefont {P.}~\bibnamefont {Haen}}, \bibinfo {author}
  {\bibfnamefont {D.}~\bibnamefont {Braithwaite}}, \bibinfo {author}
  {\bibfnamefont {G.}~\bibnamefont {Knebel}}, \bibinfo {author} {\bibfnamefont
  {S.}~\bibnamefont {Raymond}}, \ and\ \bibinfo {author} {\bibfnamefont
  {S.}~\bibnamefont {Kambe}},\ }\href {\doibase
  http://dx.doi.org/10.1016/j.jmmm.2003.11.030} {\bibfield  {journal} {\bibinfo
   {journal} {J. Magn. Magn. Mat.}\ }\textbf {\bibinfo {volume} {272-276, Part
  1}},\ \bibinfo {pages} {27 } (\bibinfo {year} {2004})}\BibitemShut {NoStop}%
\end{thebibliography}
